\documentclass[12pt]{article}

\usepackage{amsmath}
\usepackage{amssymb}
\usepackage{epsfig}

\usepackage{cite}

\usepackage{xspace}

\begin{document}

\newcommand{\half}{\frac{1}{2}}

%%%%%%%%%%%% Equation numbering:          
\renewcommand{\theequation}{\thesection.\arabic{equation}}
%%%%%%%%%%%%

\newcommand{\ope}{OPE}
\newcommand{\hwr}{highest weight representation}
\newcommand{\hws}{highest weight state}
\newcommand{\desc}{descendant}

\newcommand{\f}{\phi}
\newcommand{\p}{\psi}

\newcommand{\dd}{\partial}
\newcommand{\Vb}{V_{\vec \beta}}

\newcommand{\barray}{\setlength\arraycolsep{2pt} \begin{array}}
\newcommand{\earray}{\end{array}}
\newcommand{\dis}{\displaystyle}

\newcommand{\ketNS}[3]{ \big|\, #1 \, , #2 \, , #3 \, \big>}
\newcommand{\ket}[1]{ \big|\, #1  \, \big>}
\newcommand{\ketTW}[2]{ \big|\, #1 \, , #2 \, \big>}

\newcommand{\Rab}[4]{
\left| 
\setlength\arraycolsep{0.2ex}
\begin{array}{ccc}
#1&,& #2\\[0.4ex]
#3&,& #4\\  
\earray
\right>}

\newcommand{\NO}[2]{{:} #1 #2 {:} }
\newcommand{\NOthree}[3]{{:} #1 #2 #3 {:}}
\newcommand{\com}[2]{\left[#1,#2\right]}
\newcommand{\acom}[2]{\{#1,#2\}}

\newcommand\sww{${\mathcal{SW}}(3/2,3/2,2)$\xspace} 
\newcommand\sw{${\mathcal{SW}}(3/2,2)$\xspace}  
\newcommand\cl{ c_p^{(1)} }
\newcommand\cu{ c_p^{(2)} }

\newcommand{\aup}{a_\uparrow}
\newcommand{\adown}{a_\downarrow}
\newcommand{\bup}{b_\uparrow}
\newcommand{\bdown}{b_\downarrow}
\newcommand{\avec}[1]{{\vec \alpha}_{#1}}

\newcommand{\ii}{\mathrm{i}}
\newcommand{\ee}{\mathrm{e}}

\newcommand{\cpl}{\lambda}
\newcommand{\www}{\mu}

\newcommand{\gsub}{\widetilde G}
\newcommand{\tsub}{\widetilde T}
\newcommand{\csub}{\widetilde c}

\newcommand{\ck}[1]{c^{\scriptscriptstyle{N\!=\!1}}_{#1}}

\newcommand{\dm}[1]{d^{(#1)}}

\newcommand{\none}{$N\!=\!1$\ }

\newcommand{\sprc}{superconformal\ }
\newcommand{\sutwo}{su(2)}

\newcommand{\kt}[4]{
\rule{0cm}{28pt}
\left(
\setlength\arraycolsep{0pt}
\begin{array}{c}
#1 %\rule{0cm}{14pt}
\\[4pt]
#2,#3,#4
%\rule[-9pt]{0cm}{14pt}
\end{array}
\right)}

\newcommand{\ktR}[4]{
\rule{0cm}{28pt}
\left(
\setlength\arraycolsep{0pt}
\begin{array}{c}
{\it{#1}} %\rule{0cm}{14pt}
\\[4pt]
{\it{#2,#3,#4}}
%\rule[-9pt]{0cm}{14pt}
\end{array}
\right)}

\newcommand{\kttw}[2]{
\rule{0cm}{28pt}
\left(
\setlength\arraycolsep{0pt}
\begin{array}{c}
#1 %\rule{0cm}{14pt}
\\[4pt]
#2
%\rule[-9pt]{0cm}{14pt}
\end{array}
\right)}

\newcommand{\kttwI}[2]{
\rule{0cm}{28pt}
\left(
\setlength\arraycolsep{0pt}
\begin{array}{c}
{\it{#1}} %\rule{0cm}{14pt}
\\[4pt]
{\it{#2}}
%\rule[-9pt]{0cm}{14pt}
\end{array}
\right)}

\newcommand{\kth}[1]{#1}
\newcommand{\ktRh}[1]{\it{#1}}

\hyphenation{ho-lo-no-my ho-lo-no-mies ma-ni-fold ma-ni-folds}

%%%%%%%%%%%%%%%%%%%%%%%%%%%%%%%%%%%

\begin{titlepage}

{\hfill
\vbox{
\hbox{hep-th/0201198}
\hbox{WIS/06/02--JAN--DPP}
\hbox{January 24, 2002}
}}

\vfill

\begin{center}
{\LARGE \bf
Unitary minimal models of $\mathcal{SW}(3/2,3/2,2)$\\
superconformal algebra and\\
manifolds of $G_2$ holonomy\\
%string theory on $G_2$ manifolds\\
}
\vspace{20pt}
{
{ \large \bf Boris Noyvert\\} %\\[-2ex]
\vspace{10pt}
{  e-mail: noyvert@weizmann.ac.il\\}
\vspace{3pt}
{ \small \it Department of Particle Physics,}\\
{ \small \it  Weizmann Institute of Science,} \\
{ \small \it  76100, Rehovot,  Israel.}\\
}
%\date{}

\vspace{20pt}

\begin{abstract}

{\normalsize

The \sww superconformal algebra
is a $\mathcal{W}$ algebra with two %generic
free parameters. It consists of
3 superconformal currents of spins
$3/2$, $3/2$ and 2. The algebra is proved
to be the symmetry algebra of the coset
$\frac{su(2) \oplus su(2) \oplus su(2)}{su(2)}$.
At the central charge $c\!=\!10\half$ the
algebra coincides with the superconformal
algebra associated to manifolds of 
$G_2$ holonomy. The unitary minimal
models of the \sww algebra and their
fusion structure are found. The spectrum
of unitary representations of the
$G_2$ holonomy algebra is obtained.
}

\end{abstract}
%\maketitle
\end{center}

\vfill

\end{titlepage}

%%%%%%%%%%%%%%%%%%%%%%%%%%%%%%%%%%%
%\begin{center}
%\today
%\end{center}

\tableofcontents

%\newpage

%%%%%%%%%%%%%%%%%%%%%%%%%%%%%%%%%%%

\section{Introduction}

\setcounter{equation}{0}

%%%%%%%%%%%%%%%%%%%%%%%%%%%%%%%%%%%

Recently the manifolds of exceptional holonomy attracted
much attention. 
%in the context of the string theory
%compactifications. 
These are 7--dimensional 
manifolds of $G_2$ holonomy and 8--dimensional
manifolds of $Spin(7)$ holonomy.
They are considered in the context of the string theory
compactifications.

The supersymmetric nonlinear sigma models
on the manifolds of exceptional holonomy
are described by conformal field theories,
their superconformal chiral algebras
were constructed in \cite{Shatashvili:zw}.
We will call them the $G_2$ and $Spin(7)$
superconformal algebras. These are nonlinear 
${\mathcal{W}}$--algebras (\cite{Zamolodchikov:1985wn},
for review see \cite{Bouwknegt:1993wg}) 
of central charge
$10\half$ and $12$ respectively.
The conformal field theories were further
studied in 
\cite{
Figueroa-O'Farrill:1996hm,
gn,
Eguchi:2001xa,
Sugiyama:2001qh,
Blumenhagen:2001jb,
Roiban:2001cp,
Eguchi:2001ip,
Blumenhagen:2001qx}.

The $Spin(7)$ algebra is identified \cite{Figueroa-O'Farrill:1996hm}
with the \sw \sprc algebra 
\cite{fofs,
Komata:1991cb,
Blumenhagen:1992nm,
Blumenhagen:1992vr,
Eholzer:1992pv,
Mallwitz:1994hh}, 
existing at generic values
of the central charge.
It consists of the \none \sprc algebra 
extended by its spin--2 superprimary field.
The unitary representation theory
of the \sw algebra is studied in \cite{gn},
where complete list of unitary representations is 
determined (including the $c=12$ model, corresponding to the
$Spin(7)$ manifolds).

In this paper we identify the $G_2$
algebra with the \sww superconformal algebra
(in notations of \cite{Bouwknegt:1993wg})
at the central charge $c=10\half$ and the coupling constant
(see below) $\cpl=0$. The \sww algebra
was first constructed in \cite{Blumenhagen:1992nm}
(see also \cite{Blumenhagen:1992vr}).
It is superconformal ${\mathcal{W}}$--algebra, which 
besides the energy--momentum supercurrent
(the first ``$3/2$'' in \sww) contains 
two supercurrents of spins $3/2$ and $2$.
The \sww algebra has two generic parameters.
Along with the central charge there is 
a free coupling $\cpl$ (the self--coupling of the
spin--$3/2$ superprimary field),
which is not fixed by Jacobi identities.

In \cite{Mallwitz:1994hh} the
\sww algebra is shown to be the symmetry
algebra of the quantized Toda theory
corresponding to the $D(2|1;\alpha)$
Lie superalgebra (the only simple
Lie superalgebra with free parameter).
In the same ref.\cite{Mallwitz:1994hh} 
the free field representation of the
\sww algebra is constructed.

We study different aspects of the \sww algebra
in the present paper. First we find
that the \sww algebra is the symmetry
algebra of the diagonal coset 
\begin{equation}                              \label{coset}
  \frac{\sutwo_{k_1} \oplus \sutwo_{k_2} \oplus 
\sutwo_2}{\sutwo_{k_1+k_2 +2}} \, .
\end{equation}
We define \hwr s of the algebra and study
their unitarity. The unitary minimal
models are described by the coset (\ref{coset}).
Their central charge and coupling $\cpl$
are given by
\begin{align}
                         \label{c cpl c}
c_{k_1, k_2}&=
\frac{9}{2}  + \frac{6}{ {k_1} + {k_2}+4}
- \frac{6}{ {k_1}+2} - \frac{6}{ {k_2}+2} \, ,\\
                          \label{c cpl cpl}
\cpl_{k_1, k_2}&=
\frac{4\,{\sqrt{2}}\,\left( {k_1} - {k_2} \right) \,
    \left(  2\,{k_1} + {k_2}+6 \right) \,
    \left(  {k_1} + 2\,{k_2} +6 \right) }
{3\,{\left(3\, {k_1} \, {k_2}\, ( {k_1} + {k_2}+6) \right)^{1/2}}
    \left(  {k_1}+2 \right) 
    \left(  {k_2}+2 \right) \left(  {k_1} + {k_2}+4 \right) } \, .
\end{align} 
We also obtain all the values of $c$ and $\cpl$,
where the \sww algebra has continuous spectrum of 
unitary representations. One such model 
($c=10\half$, $\cpl=0$), which corresponds to the
$G_2$ algebra, is discussed in details,
the full spectrum of unitary representations
is obtained. We also present the complete
list of the minimal model representations
and their fusion rules.

The diagonal coset constructions 
of type 
$\frac{g \oplus g}{g}$
were found very
useful in the description of minimal models
of different conformal algebras.
The minimal models of the Virasoro algebra
\cite{BPZ}
($c_k = 1-\frac{6}{(k+2)(k+3)}$) 
correspond to the diagonal coset construction
\cite{Goddard:1985vk}
\begin{equation}
\frac{\sutwo_k \oplus \sutwo_1}{\sutwo_{k+1}} \, ,
\qquad k \in \mathbb{N} \, .
\end{equation}
The coset (\ref{coset N=1})
is found \cite{Goddard:1986ee} to form the minimal models 
of the \none superconformal algebra 
(\cite{Eichenherr:cx, Bershadsky:dq, Friedan:1984rv}
and appendix \ref{appN=1}). 
%(Their central charge 
%is given by (\ref{C_k}).)
The minimal models of the ${\mathcal{W}}_N$ algebra \cite{Lukyanov:1990tf}
are the $su(N)$ diagonal cosets
\begin{equation}
\frac{su(N)_k \oplus su(N)_1}{su(N)_{k+1}} \, ,
\qquad k \in \mathbb{N}\, .
\end{equation} 
We present here the first example (to our knowledge)
of the conformal chiral algebra, corresponding to the diagonal
coset of type $\frac{g \oplus g \oplus g}{g}$.
It is nontrivial fact that the coset space (\ref{coset})
has the same symmetry algebra for different $k_1$
and $k_2$. It can be explained, probably,
by the connection of the \sww algebra to
the Lie superalgebra $D(2|1;\alpha)$,
which has a free parameter
unlike the other simple Lie algebras.

The \sww algebra contains two fields of spin $3/2$
and three fields of spin $2$, making enough room
for embedding of different subalgebras,
such as the $N\!=\!0$ (Virasoro) and the  \none conformal
algebras. Besides the trivial \none subalgebra
(generated by the super energy--momentum tensor)
there are 3 different \none superconformal
subalgebras of the \sww algebra.
These embeddings play a crucial 
role in the understanding of the representation theory
of the algebra.  

There are four types of  \hwr s of the algebra:
Neveu--Schwarz (NS), Ramond and two twisted sectors.
(The twisted sectors are defined only in the 
case of vanishing coupling $\cpl$.)

The minimal models are labeled by two natural numbers:
$k_1$ and $k_2$. The NS and Ramond minimal model representations
can be arranged in the form of 3--dimensional
table, similarly to the 2--dimensional tables of
representations of the $N\!=\!0$ and the \none conformal
algebras. The fusion rules also satisfy the ``$\sutwo$
pattern'' of the $N\!=\!0$ and \none minimal model fusions.

The set of the $G_2$ algebra representations consists of 
4 sectors: NS, Ramond and two twisted. 
There are continuous spectrum representations
in every sector. We prove, that the $G_2$ 
conformal algebra is the extended version
of the \sw algebra at $c=10\half$. 
Due to this fact we get all the
$G_2$ unitary representations from the known
spectrum \cite{gn} of the \sw algebra.

The paper is organized as follows.
After reviewing the structure of %the
\sww 
%superconformal algebra 
in section~\ref{structure of the algebra}
we prove in section~\ref{coset constr}
that the algebra is the symmetry algebra of
the coset space (\ref{coset}).
In section~\ref{N=1 superconformal subalgebras}
we discuss different embeddings
of the \none \sprc algebra into the \sww algebra
and obtain the unitarity restrictions on the
values of $c$ and $\cpl$.
In section~\ref{Highest weight representations}
the \hwr s of the algebra are introduced, the
zero mode algebras in different sectors are discussed.
In section~\ref{minimal models} we concern 
with the minimal models of the algebra:
we explain how the spectrum of unitary representations
is obtained, discuss the fusion rules, and give two examples
of the \sww minimal models in terms of the $N\!=\!2$
\sprc minimal models. Section~\ref{G2 algebra}
is devoted to the $G_2$ algebra, the \sprc
algebra associated to the manifolds of $G_2$ holonomy.
We find it convenient to put some useful 
(but in some cases lengthy) information
in the closed form in the five appendices.

We have to note that substantial part of the
calculations was done with a help of 
{\sl Mathematica} package \cite{Thielemans:1991uw}
for symbolic computation of operator product expansions.

%%%%%%%%%%%%%%%%%%%%%%%%%%%%%%%%%%%%%%%%%%%%%%%%%%%%%%%%%%%%%

\section{Structure of the \sww algebra}

\setcounter{equation}{0}

\label{structure of the algebra}

%%%%%%%%%%%%%%%%%%%%%%%%%%%%%%%%%%%%%%%%%%%%%%%%%%%%%%%%%%%%%

Here we review the structure of \sww algebra,
which was first constructed in \cite{Blumenhagen:1992nm}.

The \sww algebra is an extension of \none superconformal 
algebra by two superconformal multiplets
of dimensions $\frac{3}{2}$ and $2$.

A superconformal multiplet $\widehat \Phi = (\Phi, \Psi)$
of dimension $\Delta$ consists of two Virasoro 
primary fields of dimensions $\Delta$ and $\Delta + \half$.
Under the action of the supersymmetry generator $G$
they transform as 
\begin{align}
G(z)\,\Phi(w)&= \frac{\Psi(w)}{z-w}   \, ,   \label{GPhi}   \\
G(z)\,\Psi(w)&=\frac{2\, \Delta \, \Phi(w)}{(z-w)^2}+
\frac{\dd \Phi(w)}{z-w}  \, .
\end{align}

The \sww algebra multiplets are denoted by 
$I=(G, T)$, $\widehat H = (H,M)$ ($\Delta = \frac{3}{2}$),
$\widehat W = (W,U)$ ($\Delta = 2$). 
The structure of the algebra is schematically given by 
\begin{equation}                               \label{algebra_structure}
\begin{aligned}
\widehat H \times \widehat H &= 
I + \cpl \, \widehat H + \www \, \widehat W \, ,\\
\widehat H \times \widehat W &= 
\www \, \widehat H+ \cpl \, \widehat W \, ,\\
\widehat W \times \widehat W &= 
I + \cpl \, \widehat H + \www \, \widehat W 
+ %1/c \, 
\NO{\widehat H}{\widehat H} \, ,
\end{aligned}
\end{equation}
where 
\begin{equation}
  \www = \sqrt{\frac{9\, c\, (4+\cpl^2)}{2 \, (27-2\, c)}} \, ,
\end{equation}
and the $c$ dependence of the coefficients is omitted.

The explicit operator product expansions
(\ope s) are fixed by the fusions
(\ref{algebra_structure})
and by \none superconformal invariance. We reproduce
the \ope s in appendix~\ref{apA}.

Unitarity is introduced by the standard conjugation relation 
$\mathcal{O}_n^\dagger = \mathcal{O}_{-n}$ for any generator 
except $U$.  The commutation relation
$\com{G_n}{W_m} = U_{n+m}\, $, following 
from (\ref{GPhi}), requires the conjugation 
$U_n^\dagger = -U_{-n}$.

%%%%%%%%%%%%%%%%%%%%%%%%%%%%%%%%%%%%%%%%%%%%%%%%%%%%%%%

\section{Coset construction}

\setcounter{equation}{0}

\label{coset constr}

%%%%%%%%%%%%%%%%%%%%%%%%%%%%%%%%%%%%%%%%%%%%%%%%%%%%%%%

\subsection{Preliminary discussion}

%%%%%%%%%%%%%%%%%%%%%%%%%%%%%%%%%%%%%%%%%%%%%%%%%%%%%%%

We start from a few supporting arguments,
that the coset theory (\ref{coset}) indeed possesses
the \sww superconformal symmetry.

First, we note that formally $\sutwo_2 \approx so(3)_1$
and the coset (\ref{coset}) can be written in the form
of Kazama--Suzuki coset \cite{Kazama:1989qp}
$\dis \frac{g \oplus so(\dim{g}-\dim{h})_1}{h}$,
where $g= \sutwo \oplus \sutwo$ and $h$ is its $\sutwo$
diagonal subalgebra. It means, that the chiral algebra 
contains \none superconformal algebra, obtained 
from the affine currents by a superconformal generalization
of Sugawara construction (see ref.~\cite{Kazama:1989qp}
for details).

All other currents, that constitute the chiral algebra
should come in pairs of superpartners with difference
of scaling dimensions equal to $1/2$.

The central charge (\ref{c cpl c}) of the coset models (\ref{coset})
%\begin{equation}                          \label{eq:central_charge}
% c_{k_1, k_2}=\frac{3\, k_1}{k_1+2}+
%\frac{3\, k_2}{k_2+2}+\frac{3}{2}-
%\frac{3\, (k_1+k_2+2)}{k_1+k_2+4}
%\end{equation}
has limiting point (when $k_1, k_2 \to \infty $)
$c=9/2$.
All the known examples of minimal series have 
limiting central charge $c=n_B+\half n_F$, where
$n_B$ and $n_F$ are the number of bosonic and fermionic
fields in the correspondent chiral algebra. 
Adopting the argument to our case we get,
that the chiral algebra consists of 
three supercurrents (including the super--Virasoro
operator).

The next argument follows from the simple
observation \cite{Kazama:1989uz}, 
that if there is a sequence
of subalgebra inclusions 
\begin{equation}
  g \supset h_1 \supset \ldots \supset h_n
\end{equation}
then the coset theory can be decomposed to the %tensor product
direct sum
\begin{equation}
  \frac{g}{h_n} = \frac{g}{h_1} \oplus \frac{h_1}{h_2}
\oplus \ldots \oplus \frac{h_{n-1}}{h_n}\, .
\end{equation}
In the case of coset (\ref{coset}) the inclusion sequence is
\begin{equation}
  \sutwo_{k_1} \oplus \sutwo_{k_2} \oplus \sutwo_2
\supset h_1 \supset \sutwo_{k_1+k_2+2}
\end{equation}
with 3 different choices of $h_1$:
$\sutwo_{k_1+2} \oplus \sutwo_{k_2}\, ,$ \\
$\sutwo_{k_2+2} \oplus \sutwo_{k_1}\, ,$
$\sutwo_{k_1+k_2} \oplus \sutwo_2\, .$
%
%\begin{align}
%\sutwo_{k_1+k_2} &\oplus \sutwo_2 \\
%\sutwo_{k_1+2} &\oplus \sutwo_{k_2} \\
%\sutwo_{k_2+2} &\oplus \sutwo_{k_1}
%\end{align}
%
The correspondent decompositions are:
\begin{align}
\label{decom2}
\frac{\sutwo_{k_1} \oplus \sutwo_{2}}
{\sutwo_{k_1+2}} 
&\oplus 
\frac{\sutwo_{k_1+2} \oplus \sutwo_{k_2}}
{\sutwo_{k_1+k_2+2}} \, ,\\
\label{decom3}
\frac{\sutwo_{k_2} \oplus \sutwo_{2}}
{\sutwo_{k_2+2}} 
&\oplus 
\frac{\sutwo_{k_2+2} \oplus \sutwo_{k_1}}
{\sutwo_{k_1+k_2+2}} \, ,\\
\label{decom1}
\frac{\sutwo_{k_1} \oplus \sutwo_{k_2}}
{\sutwo_{k_1+k_2}} 
&\oplus 
\frac{\sutwo_{k_1+k_2} \oplus \sutwo_2}
{\sutwo_{k_1+k_2+2}} \, .
\end{align}
All three contain coset spaces of type (\ref{coset N=1})
with  $k=k_1$, $k=k_2$ and $k=k_1+k_2$ respectively.
Therefore the chiral algebra contain 3 different
\none superconformal subalgebras (not including
the trivial one, generated by the 
super energy--momentum tensor).
This is possible if there are at least 3 operators
of scaling dimension 2. They have 3 superpartners 
of dimensions $\frac{3}{2}$ or $\frac{5}{2}$.
One field of dimension $\frac{3}{2}$ can not serve
as a superconformal generator for 3 different superconformal
subalgebras. The case, when all three are of dimension
$\frac{3}{2}$ is also excluded, because then the algebra 
is trivially a sum of 3 commuting \none superconformal
algebras.

Collecting the arguments we get that the 
only possibility that the chiral algebra 
consists of 6 fields of dimensions 
$\frac{3}{2}$, $\frac{3}{2}$, 2, 2, 2, $\frac{5}{2}$,
which can be combined to three supercurrents 
of dimensions $\frac{3}{2}$, $\frac{3}{2}$, 2,
giving the \sww algebra.

%%%%%%%%%%%%%%%%%%%%%%%%%%%%%%%%%%%%%%%%%%%%%%%%%%%%%%%%%%%%%

\subsection{Explicit construction}

\label{sec:expli_const}

%%%%%%%%%%%%%%%%%%%%%%%%%%%%%%%%%%%%%%%%%%%%%%%%%%%%%%%%%%%%%

In this section we prove by explicit construction
that the coset (\ref{coset}) contains
the \sww algebra. The method we use was 
first proposed in \cite{Goddard:1986ee} for coset (\ref{coset N=1}).

The $\sutwo$ affine algebra is generated by 3 currents
$J_i$, $i=1,2,3$:
\begin{equation}
  J_i (z) \, J_i (w) = \frac{k/2}{(z-w)^2}+ 
\frac{\ii \, \epsilon_{i j k} \, J_k (w)}{z-w} \, .
\end{equation}
The ${g}$ algebra consists of 3 commuting 
copies of the $su(2)$ algebra at levels 
$k_1$, $k_2$ and $2$. The $\sutwo$ on level $2$
is realized by free fermions in the adjoint representation of
$\sutwo$:
\begin{equation}
  \psi_i (z) \, \psi_j (w)=\frac{\delta_{i j}}{z-w} \, ,
\qquad i,j = 1,2,3.
\end{equation}
Then the affine currents are expressed as 
\begin{equation}
  J_i = - \frac{\ii}{2}  \, \epsilon_{i j k} \, 
\NO{\psi_j}{\psi_k} \, .
\end{equation}
The affine algebra ${h} =\sutwo_{k_1+k_2+2}$ is diagonally
embedded in 
${g}=\sutwo_{k_1} \oplus \sutwo_{k_2} \oplus \sutwo_2$:
\begin{equation}
  J_i^{(h)}=J_i^{(1)}+J_i^{(2)}+J_i^{(3)} \, .
\end{equation}
The coset space ${g}/{h}$ contains operators,
constructed from the ${g}$ currents, which commute
with the currents of ${h}$.
The energy--momentum tensor of the coset ${g}/{h}$
is given by the Sugawara construction:
\begin{equation}
  T=T^{(g)}-T^{(h)}=T^{(1)}+T^{(2)}+T^{(3)}-T^{(h)}\, ,
\end{equation}
where
\begin{equation}
  T^{(n)}=\frac{1}{k_n+2}
\sum_{i=1}^3 \NO{J_i^{(n)}}{J_i^{(n)}} \, .
\end{equation}
%% and then the central charge is as in (\ref{c cpl}).

The general dimension--$3/2$ operator can be written as
\begin{equation}
  \mathcal{O}_{3/2} =
b_1 \sum_{i=1}^3 \NO{J_i^{(1)}}{\psi_i} +
b_2 \sum_{i=1}^3 \NO{J_i^{(2)}}{\psi_i} +
\ii \, b_3 \, \NOthree{\psi_1}{\psi_2}{\psi_3}\, .
\end{equation}
%The operators, which belong to the coset space ${g}/{h}$
It should commute with the $J^{(h)}$ currents.
This requirement leads to condition
\begin{equation}
b_3 = \half (b_1 \, k_1 +b_2 \, k_2) \, .
\end{equation}
One has two independent dimension--$3/2$ operators ($G$ and $H$)
in the coset theory, since there are two free parameters 
$b_1$ and $b_2$. 
%We choose one of the operators ($G$)
%to be the generator
%of the \none superconformal symmetry,
%setting
%\begin{equation}
%  \begin{split}
%          b_1(2+k_1)+b_2(2+k_2)=0\, ,\\
%          b_1=\sqrt{\frac{2\, (k_2+2)}{(k_1+2)(k_1+k_2+4)}} \, .
%  \end{split}
%\end{equation}
%The second operator H is chosen to be the \none superconformal
%primary field, its \ope\  with $G$ should start from the first
%order pole term:
%\begin{equation}
%G(z)\, H(w)=\frac{M(w)}{z-w} \, ,
%\end{equation}
%leading to condition
%\begin{equation}
%b_1 \, k_1 \, (k_1+2\, k_2+6) = 
%b_2 \, k_2 \, (2\, k_1+k_2+6)\, ,
%\end{equation}
%and the convenient normalization is 
%\begin{equation}
%b_1= -2\, (2\, k_1+k_2+6) 
%\sqrt{\frac{k_2 / k_1}{(k_1+2)(k_2+2)(k_1+k_2+4)(k_1+k_2+6)}} \, .
%\end{equation}

In order to close the algebra one needs 3 more operators 
$M$, $W$, $U$ of scaling dimensions 2, 2, $5/2$.
The coset construction of all the 6 operators is given 
in appendix~\ref{apB}.

We have explicitly checked, that the set of 6 operators
$T$, $G$, $H$, $M$, $W$, $U$ satisfy the \ope s of
the \sww algebra 
%(see the next section and appendix~\ref{apA}) 
with central charge $c$
and 
coupling $\cpl$ given by (\ref{c cpl cpl}).

%%%%%%%%%%%%%%%%%%%%%%%%%%%%%%%%%%%%%%%%%%%%%%%%%%%%%%%%%%%%%%

\section{\none superconformal subalgebras}

\setcounter{equation}{0}

\label{N=1 superconformal subalgebras}

%%%%%%%%%%%%%%%%%%%%%%%%%%%%%%%%%%%%%%%%%%%%%%%%%%%%%%%%%%%%%%

We start this section by observation, that 3 bosonic 
operators $T, M, W$ do not generate a closed subalgebra
because of $\NO{G}{H}$ term in the \ope\ of $M$ with $W$
(\ref{opeMW}).

We will construct in this section different \none 
superconformal subalgebras of \sww and discuss
the unitarity restrictions. The most general 
dimension--$\frac{3}{2}$ operator is 
\footnote{
The discussion applies to NS and Ramond sectors.
One cannot mix $G$ and $H$ in the case of twisted
sectors.
} 
\begin{equation}
  \gsub =\alpha \, G + \beta \, H \, .
\end{equation}
We calculate its \ope\  with itself:
\begin{equation}
  \gsub(z) \, \gsub(w) = 
\frac{\frac{2}{3} \, c \, (\alpha^2 + \beta^2)}{(z-w)^3}+
\frac{2\, \tsub}{z-w} \, ,
\end{equation}
where 
\begin{equation}
\tsub = (\alpha^2 +\beta^2) \, T + 
\beta \, (\alpha + \half \, \cpl \, \beta) \, M +
\frac{2}{3} \, \www \, \beta^2 \, W \, ,
\end{equation}
and take the dimension--2 operator $\tsub$
as the Virasoro generator of the subalgebra.
These two operators, $\gsub$ and $\tsub$,
generate a closed subalgebra if the following
two equations are satisfied:
\begin{equation}          \label{eq:alpha_beta}
\begin{aligned}
%\beta \left (
27 \, \alpha^2 + 27 \, \cpl \, \alpha \, \beta +
( 9 \, \cpl^2 +4 \, \www^2 + 9) \, \beta^2 -9 
%\right) 
&= 0 \, ,\\
\alpha^3 - \alpha +3 \, \alpha \, \beta^2 + \cpl \, \beta^3 &= 0 \, .
\end{aligned}
\end{equation}
Then the central charge of the subalgebra is 
$\csub = c\, (\alpha^2+\beta^2)$.
Formally the lefthand side of the first equation 
should be multiplied by $\beta$. We removed it in order
to exclude the trivial solution $\beta=0, \alpha=1$,
corresponding to the obvious \none subalgebra.

The operator $T-\tsub$ generates
another closed subalgebra, namely Virasoro
algebra with central charge $c-\csub\,$, and it is commutative
with the \none superconformal subalgebra of $\gsub$ and $\tsub$.

The equations (\ref{eq:alpha_beta}) are polynomial
equations in $\alpha$ and $\beta$ of orders 2 and 3.
Generically there are 6 solutions. One should take
into account the $\mathbb{Z}_2$ symmetry 
$\alpha \to - \alpha$ and $\beta \to - \beta$
of the equations, corresponding to the $\mathbb{Z}_2$
transformation of the \none superconformal algebra
$\gsub \to - \gsub, \,\tsub \to  \tsub $. So
at any generic $c$ and $\cpl$ there are three 
\none superconformal subalgebras. 

However, we are interested in real subalgebras, i.e. 
preserving the conjugation relation:
the conjugation of the subalgebra 
$\gsub_n^\dagger = \gsub_{-n}, \, \tsub_n^\dagger = \tsub_{-n}$
should be consistent with the conjugation
relations of the \sww algebra.
This is true, if $\alpha$ and $\beta$ are real.
Generically there can be 1 or 3 real solutions 
of~(\ref{eq:alpha_beta}).

Unitary representations of an algebra are necessarily 
unitary representations of all
its real subalgebras. The \none superconformal algebra
has unitary representations only at
$\csub \ge 3/2$ or at
$\csub= \ck{k}\, , \, k \in \mathbb{N}\,$ (\ref{C_k}).
%\frac{3}{2}-\frac{12}{(k+2)(k+4)}\, , \, k \in \mathbb{N}$
%(ref.~\cite{N1unitarity}).
The central charges of all the real
\none superconformal subalgebras should be from this set.

We study the solutions of the set of equations (\ref{eq:alpha_beta})
in the region $0 \le c < \frac{27}{2}\,$.
(At $c > \frac{27}{2}$ the coupling $\www$ becomes
imaginary.) The results are presented in figure~\ref{figure1}.
The region $0 \le c < \frac{27}{2}$ is divided to two parts
by the curve (thick curve in figure~\ref{figure1})
\begin{equation}   \label{divide curve}
4\, (9-2\, c)^3 = 243\, (2\, c-3) \, \cpl^2\, .
\end{equation}
\begin{figure}                          
\centering
\includegraphics[width=0.95 \textwidth]{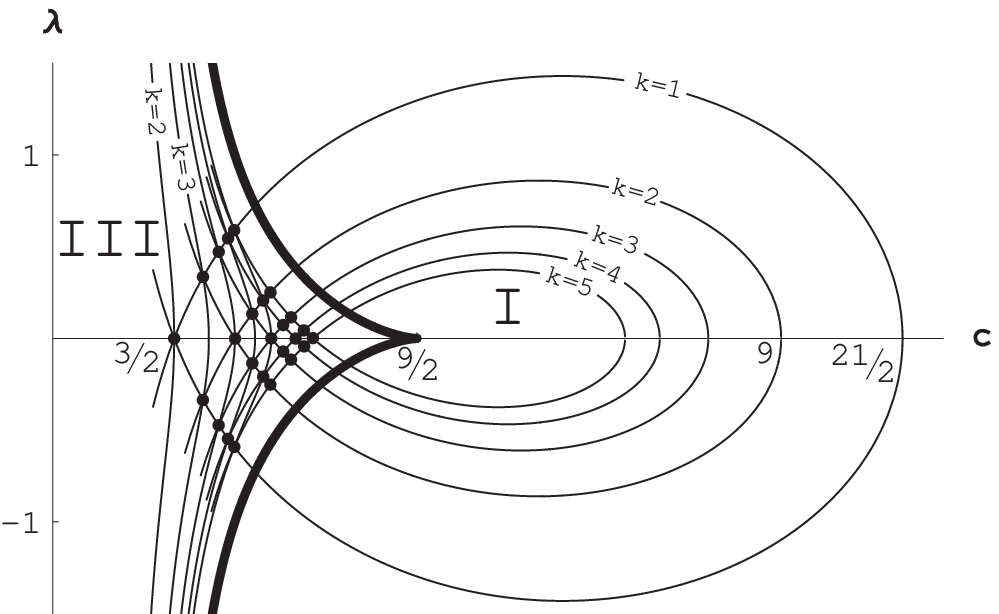}
{\caption{Unitary models of the \sww algebra.}   \label{figure1}}
\end{figure}
In the region I
($4\, (9-2\, c)^3 < 243\, (2\, c-3) \, \cpl^2$)
there is one real solution of (\ref{eq:alpha_beta}),
in the region III 
($4\, (9-2\, c)^3 > 243\, (2\, c-3) \, \cpl^2$)
there are 3 real solutions.

The curves in figure~\ref{figure1} have constant subalgebra 
central charge along it: $\csub=\ck{k}$. We call them
the unitary curves.
Taking different 
\none subalgebras one gets different curves:
\begin{align} 
              \label{curveA}
{\cpl}^2 &=
\frac{4\,\left( 54 - 4\,c + 9\,k - 2\,c\,k \right) \,
  {\left( 4\,c - 9\,k + 2\,c\,k \right) }^2}
{243\,k\,
  \left( 16\,c - 18\,k + 12\,c\,k - 3\,k^2 + 
    2\,c\,k^2 \right)} \\
{\cpl}^2 &=        \label{curveB}
\frac{4\,\left( 9\,k - 2\,c\,k -8\,c \right)
      \,{\left(  8\,c  + 
        2\,c\,k - 9\,k -54 \right) }^2}{243\,
    \left( 6 + k \right) \,
    \left( 16\,c - 18\,k + 12\,c\,k - 3\,k^2 + 
      2\,c\,k^2 \right) }
\end{align}
%\qquad  k \in \mathbb{N} 
%k=1,2,3, \ldots
%are the curves of constant subalgebra central charge
%$\csub=\ck{k}\,$. 
All the region under discussion is spanned
by the curves (\ref{curveA}, \ref{curveB}).
There are no real solutions of (\ref{eq:alpha_beta})
corresponding
to $\csub > 3/2\,$.

In the region III
the unitarity is restricted to the intersections 
of the unitary curves (the dots in figure~\ref{figure1}). 
There are intersections of exactly 3 curves in 
every intersection point: two curves of type
(\ref{curveA}) with $k=k_1$ and $k=k_2$ and the third
of type (\ref{curveB}) with $k=k_1+k_2$.
The intersection points are given by (\ref{c cpl c}, \ref{c cpl cpl}).
%\begin{equation}                          \label{c cpl}
%\begin{aligned}
%c_{k_1, k_2}&=
%\frac{9}{2}  + \frac{6}{4 + {k_1} + {k_2}}
%- \frac{6}{2 + {k_1}} - \frac{6}{2 + {k_2}}\\
%%
%\cpl_{k_1, k_2}&=
%\frac{4\,{\sqrt{2}}\,\left( {k_1} - {k_2} \right) \,
%    \left( 6 + 2\,{k_1} + {k_2} \right) \,
%    \left( 6 + {k_1} + 2\,{k_2} \right) }
%{3\,{\sqrt{3\, {k_1} \, {k_2}\, (6 + {k_1} + {k_2})}}
%    \left( 2 + {k_1} \right) 
%    \left( 2 + {k_2} \right) \left( 4 + {k_1} + {k_2} \right) }
%\end{aligned}
%\quad  k_1 , k_2 \in \mathbb{N}
%%k_1 , k_2 =1,2, \ldots
%\end{equation}
The formula (\ref{c cpl c}) is exactly the formula for central 
charge of coset theories (\ref{coset}). 
The central charges of three real subalgebras 
(we will call them the first, the second and
the third \none subalgebras) 
at $c$ and $\cpl$  at the intersection point are
$\ck{k_1}$, $\ck{k_2}$, $\ck{k_1 + k_2}$, they also coincide with
the central charges of the three \none subalgebras 
of the coset (\ref{coset}) (see the decompositions
(\ref{decom2}, \ref{decom3}, \ref{decom1})).
We conclude that all unitary models of
the \sww algebra in the region III are given
by coset models  (\ref{coset}).
One can solve the equations~(\ref{eq:alpha_beta}) 
for the values of $c$ and $\cpl$ from (\ref{c cpl c}, \ref{c cpl cpl})
to get the linear connection between the generators
$T$, $M$, $W$ and the Virasoro generators of the 
three subalgebras:
{\small
\begin{align}     
\label{h_to_ddd}
T&=\frac{1}{2}\,\left( \tsub^{(1)}\,
      \left({k_1} + 4 \right)  + 
     \tsub^{(2)}\,\left({k_2} + 4 \right)  - 
     \tsub^{(3)}\,\left({k_1} +{k_2} + 2 \right) 
     \right),
\\
\label{m_to_ddd}
M&=
\frac{1}{\left( 6\, k_1 \, k_2 \, (k_1+k_2+6) \right)^{1/2}}
\times \nonumber \\
&\quad \times
\Bigg(
\tsub^{(3)}\,
\frac{(k_1-k_2)(k_1+k_2+2)(k_1 \, k_2-2\, k_1-2\,k_2-12)}
{(k_1+2)(k_2+2)} \Bigg. \nonumber \\
&\quad - \Bigg.
\left(
\tsub^{(1)}\,
\frac{(k_1+4)(k_1+2\,k_2+6)(k_2^2+k_1\,k_2-2\,k_1+6\,k_2)}
{(k_2+2)(k_1+k_2+4)}
-\left(1 \leftrightarrow 2 \right)
\right)
\Bigg),
\\
\label{w_to_ddd}
W&=
\left(
\frac{
3\,{{k_1}}^2\,{k_2} + 
   3\,{k_1}\,{{k_2}}^2 + 2\,{{k_1}}^2 + 
   2\,{{k_2}}^2 + 20\,{k_1}\,{k_2} + 
   12\,{k_1} + 12\,{k_2}}
{24\, k_1 \, k_2 \, (k_1+k_2+6)}
\right)^{1/2}
\times \nonumber \\
&\quad \times
\Bigg(
\tsub^{(3)}\,
\frac{(k_1+k_2+2)(k_1 \, k_2+4\, k_1+4\,k_2+12)}
{(k_1+2)(k_2+2)} \Bigg. \nonumber \\
&\quad - \Bigg.
\left(
\tsub^{(1)}\,
\frac{(k_1+4)(k_2^2+k_1\,k_2+4\,k_1+6\,k_2+12)}
{(k_2+2)(k_1+k_2+4)}
+\left(1 \leftrightarrow 2 \right)
\right)
\Bigg).
\end{align}
}

In the region I the unitarity is restricted 
to the unitary curves (\ref{curveA}). 
The region I models with $c$ and $\cpl$ satisfying (\ref{curveA})
are expected to have continuous spectrum of
unitary representations. 

On the separating curve (\ref{divide curve}) the unitarity is restricted
to the limiting points of (\ref{c cpl c}, \ref{c cpl cpl}) at one $k$ fixed
and another $k$ taken to %$\infty$. 
infinity. (And $c=9/2, \, \cpl=0$, when 
both $k_1, k_2 \to \infty$.)

The Virasoro subalgebras, generated by $T-\tsub$,
give no new restrictions on unitarity of the \sww algebra.

The point $c=10\half$, $\cpl=0$, which corresponds to
the conformal algebra on $G_2$ manifold, is in the region I
and lies on the (\ref{curveA}) curve with $k=1$. It means
that the algebra have one real \none subalgebra and its central charge
is $\ck{1}=7/10$ in agreement with results of ref.\cite{Shatashvili:zw}.
The generators of this real subalgebra are 
\begin{align}
\label{gsub_cpl_zero}
\gsub&=\left(\frac{27-2c}{3\,(9-2\,c)}\right)^{1/2} \! H ,\\
\label{tsub_cpl_zero}
\tsub&=\frac{1}{3\,(9+2\,c)}
\left(\left( 27 - 2\,c \right) \,
    T + 2\,{\left( 2\,c\,
        \left( 27 - 2\,c \right)  \right) }^{1/2}
     \,W \right).
\end{align}
(This is true for any $\cpl=0$ model.)

The important question for understanding
the structure of unitary representations of the \sww algebra
is how the algebra 
is decomposed to the representations of its real
\none subalgebras? The decomposition is 
$\Phi_{1 1} + \Phi_{3 1}$ under subalgebras 
corresponding to the (\ref{curveA}) curve 
and $\Phi_{1 1} + \Phi_{1 3}$ under subalgebras 
corresponding to the (\ref{curveB}) curve,
where $\Phi_{1 1}$ is the vacuum representation
of the \none \sprc algebra, 
$\Phi_{1 3}$ and  $\Phi_{3 1}$ are its
degenerate representations, having null vector
on level $3/2$.
The $\tsub$, $\gsub$ and $T$ fields are in the
$\Phi_{1 1}$ representation. Three other fields 
of \sww form $\Phi_{3 1}$ (or $\Phi_{1 3}$) representation,
they can be understood in this context as 
$\Phi_{3 1}$ (spin--$3/2$ field), $\gsub_{-1/2} \Phi_{3 1}$
(spin--$2$ field) and $\tsub_{-1} \Phi_{3 1}$
(spin--$5/2$ field). $\tsub_{-1} \gsub_{-1/2} \Phi_{3 1}$
is proportional to $\gsub_{-3/2} \Phi_{3 1}$ 
($\approx \NO{\gsub}{\Phi_{3 1}}$), since there is a null
state on level $3/2$. 
%of $\Phi_{3 1}$ ($\Phi_{1 3}$)
%NS representation of \none superconformal algebra.

%The decomposition is very important for understanding
%of the structure of unitary representations of the \sww algebra.

%%%%%%%%%%%%%%%%%%%%%%%%%%%%%%%%%%%%%%%%%%%%%%%%%%%%%%%%%%%%%%%%

\section{Highest weight representations}

\setcounter{equation}{0}

\label{Highest weight representations}

%%%%%%%%%%%%%%%%%%%%%%%%%%%%%%%%%%%%%%%%%%%%%%%%%%%%%%%%%%%%%%%%

The \sww commutation relations admit two consistent choices
of generator modes in the general case:
NS and Ramond sectors;
 and 
two more  then the coupling
$\cpl=0$: first twisted (tw1) and second twisted sectors (tw2).
\begin{description}

\item[NS sector.]
The modes of the bosonic operators ($L_n$, $M_n$, $W_n$) are
integer ($n \in \mathbb{Z}$)
and the modes of the fermionic operators ($G_r$, $H_r$, $U_r$)
are half--integer ($r \in \mathbb{Z}+\half$).

\item[Ramond sector.]
The modes of all the operators are integer.

\item[First twisted sector.]
The modes of  $L_n$, $W_n$, $H_n$ operators are
integer ($n \in \mathbb{Z}$)
and the modes of $G_r$, $M_r$, $U_r$ operators
are half--integer ($r \in \mathbb{Z}+\half$).

\item[Second twisted sector.]
The modes of   $L_n$, $G_n$, $W_n$, $U_n$ operators are
integer ($n \in \mathbb{Z}$)
and the modes of $H_r$, $M_r$  operators
are half--integer ($r \in \mathbb{Z}+\half$).

\end{description}

How can one understand the existence of four 
different sectors in terms of the coset construction
(appendix~\ref{apB})? 
In order to get NS or Ramond sectors
of the algebra one should take all the three fermions
of $\sutwo_2$ in NS or Ramond sectors respectively.
The modes of $\sutwo_{k_1}$ and $\sutwo_{k_2}$
currents are integer.
The twisted sectors ($k_1=k_2$, since $\cpl=0$) are obtained in
less obvious way. 
First twisted sector:
one takes 
one Ramond fermion (say, $\psi_1$) and two NS fermions
($\psi_2$ and $\psi_3$), the modes of 
$(J_1^{(1)}+J_1^{(2)})$, $(J_2^{(1)}-J_2^{(2)})$,
 $(J_3^{(1)}-J_3^{(2)})$ are integer and 
the modes of 
$(J_1^{(1)}-J_1^{(2)})$, $(J_2^{(1)}+J_2^{(2)})$,
 $(J_3^{(1)}+J_3^{(2)})$ are half--integer.
Second twisted sector: one NS fermion ($\psi_3$)
and two Ramond fermions ($\psi_1$ and $\psi_2$),
the modes of 
$(J_1^{(1)}-J_1^{(2)})$, $(J_2^{(1)}-J_2^{(2)})$,
 $(J_3^{(1)}+J_3^{(2)})$ are integer
and the modes of 
$(J_1^{(1)}+J_1^{(2)})$, $(J_2^{(1)}+J_2^{(2)})$,
 $(J_3^{(1)}-J_3^{(2)})$ are half--integer.
One cannot define the modes of separate bosonic
currents (e.g. $J_1^{(1)}$) in the twisted sectors.

% We want to make a general remark on mode
% expansions of normal ordered products of
% operators in various sectors. The well known
% formula for the mode expansion of composite
% operator
% \begin{equation}                      
%   \NO{P}{Q}_n =\sum_{m \le -\Delta_P} P_m Q_{n-m}+
% (-1)^{PQ} \! \! \! \! \! \sum_{m \ge -\Delta_P +1} Q_{n-m} P_m \, ,
% \end{equation} 
% works only if $P$ is bosonic and $m$ is integer
% or if $P$ is fermionic and $m$ is half--integer.
% Other cases should be treated with some care,
% as it is explained in details in appendix~C of ref.~\cite{gn}.
% Some nontrivial composite operator expansions in different sectors
% are used below, but for brevity
% we don't reproduce them in this publication.

The commutation relations of the \sww algebra include
products of the generators. The formula (\ref{NOepsilon})
for the
mode expansion of composite operators in various 
sectors is derived in appendix~\ref{NOexp}.

Now are ready to define the \hwr s in all sectors.
The \hws\ is annihilated by positive modes of all generators:
\begin{equation}
  \mathcal{O}_n \ket{\text{hws}}=0\, , \qquad n > 0 \, .
\end{equation}
To deal with the zero modes one should discuss the sectors separately.

%%%%%%%%%%%%%%%%%%%%%%%%%%%%%%%%%%%%%%%%%%%%%%%%%%%%%%%%%%%%%%%%

\subsection{NS sector}

%%%%%%%%%%%%%%%%%%%%%%%%%%%%%%%%%%%%%%%%%%%%%%%%%%%%%%%%%%%%%%%%

There are 3 zero modes: $L_0$, $M_0$, $W_0$.
It is convenient to choose the \hws\ to be the eigenstate
of these 3 operators and to label the  
\hwr\ by the correspondent eigenvalues.
This is possible if the set of zero modes is
commutative. One finds from (\ref{opeMW}) that the
commutator
\begin{equation}
  \com{M_0}{W_0} = \frac{9 \, \www}{2\, c}
(M_0+ \NO{G}{H}_0)
\end{equation}
is not zero.
We rewrite the commutator by expanding $\NO{G}{H}_0$ in the modes of 
$G$ and $H$:
\begin{equation}             \label{comMW}
 \com{M_0}{W_0} = \frac{9 \, \www}{2\, c} \sum_{r=1/2}^\infty 
\left( G_{-r} \, H_r -  H_{-r} \, G_r \right)
\end{equation}
% (See appendix C of ref.~\cite{gn}
%for the definition of mode expansion of normal ordered
%products of operators in various sectors.)
The action of righthand side of (\ref{comMW}) on \hws\
vanishes. This is what one effectively needs
in order to choose the \hws\ to be the eigenstate
of both $M_0$ and $W_0$. We define the notion of ``effective''
commutator: the commutation relation, which is true
modulo terms, their action on \hws\ is zero.
Concluding: the ``effective'' commutators of all
three zero modes vanish, and one can label the
\hwr\ by three weights, the eigenvalues
of the zero modes:
\begin{equation}
\begin{aligned}
L_0 \, &\ketNS{h}{m}{w} &= h &\ketNS{h}{m}{w} \, ,\\
M_0 \, &\ketNS{h}{m}{w} &= m &\ketNS{h}{m}{w} \, ,\\
W_0 \, &\ketNS{h}{m}{w} &= w &\ketNS{h}{m}{w} \, .
\end{aligned}
\end{equation}
One gets all states in the representation
acting by negative modes on the \hws .

%%%%%%%%%%%%%%%%%%%%%%%%%%%%%%%%%%%%%%%%%%%%%%%%%%%%%%%%%%%%%%%%

\subsection{Ramond sector}

%%%%%%%%%%%%%%%%%%%%%%%%%%%%%%%%%%%%%%%%%%%%%%%%%%%%%%%%%%%%%%%%

There are 6 zero modes: $L_0$, $M_0$, $W_0$, $G_0$, $H_0$, $U_0$.
Since $L_0$ is commutative with all other zero modes,
it can be represented by a number $h$.
%\begin{equation}
%  L_0 \, \ket{h} =h \ket{h}
%\end{equation}
The  (anti)commutation relations of other zero modes are
%given in appendix~\ref{zemoal} in (\ref{zmaGG}--\ref{zmaUU}). 
\begin{align}
\acom{G_0}{G_0}         \label{zmaGG}
&=  2\, (h - c/24)  \, ,\\ %(h -\frac{c}{24})  \, ,\\ 
\acom{G_0}{H_0}
&=M_0 \, ,\\ 
\com{G_0}{M_0}
&=\acom{G_0}{U_0}=0\, ,\\ 
\com{G_0}{W_0}
&=U_0 \, ,\\ 
\acom{H_0}{H_0}       \label{zmaHH}
&=  2\, (h - c/24)+ 
\cpl \, M_0 + 4/3 \, \www  \, W_0 \, ,\\ 
\com{H_0}{M_0}        \label{zmaHM}
&=  2/3 \, \www \,  U_0 \, ,\\ 
\com{H_0}{W_0}       \label{zmaHW}
&=  \cpl /2 \, U_0 \, ,\\ 
\com{M_0}{W_0}       \label{zmaMW}
&=  \acom{H_0}{U_0}=
\frac{9 \, \www}{4\, c} \, (2\, G_0 \, H_0 - M_0) \, ,\\ 
\com{M_0}{U_0}
&= \frac{9 \, \www}{2\, c} \, 
\left( 2\,(h-c/24)\, H_0 -G_0 \, M_0 \right) \, ,\\ 
\com{W_0}{U_0}
&= \frac{1}{8\,c}\, \Big(
54 \, \www \, U_0 
- 54  \,  (h - c/24) \, \left( 2\, G_0+ \cpl \, H_0
\right)\Big. \nonumber \\
& \quad \Big. -27 \, \cpl \, G_0 \, M_0 - 72 \, \www \, G_0 \, W_0 
+54 \, H_0 \, M_0 \Big) \, ,\\ 
\acom{U_0}{U_0}              \label{zmaUU}
&= \frac{9}{4\,c} \, \Big(
-12\, (h - c/24)^2-
  (h - c/24) \left(6\, \cpl \,M_0
+8\,\www\, W_0 \right) \Big. \nonumber \\
&\quad \Big. +4 \, \www \, G_0 \, U_0
+3\, M_0 \, M_0 \Big)\, .
\end{align}
They define finite dimensional $\mathcal{W}$ superalgebra.
The commutation relations (\ref{zmaGG}--\ref{zmaHW}) are exact,
and the commutators (\ref{zmaMW}--\ref{zmaUU}) are ``effective'',
i.e. modulo terms which annihilate \hws s.

The irreducible representations of the zero mode algebra 
are one--dimensional or two--dimensional and labeled by three 
weights: $h$, $w$ and $m$.
 In one--dimensional
representation the zero modes are given by
\begin{equation}
\begin{aligned}
  W_0&=w, \qquad M_0=m, \qquad U_0=0,\\
 G_0&=\sqrt{h-c/24}\, ,\quad 
H_0=\frac{m}{2\,\sqrt{h-c/24}}\, .
\end{aligned}
\end{equation}
Such a representation exists only if the following
condition is satisfied:
\begin{equation}          \label{eqR1}
  12\, (h-c/24)^2 +6\, \cpl\, (h-c/24)\, m+
8\, \www \,  (h-c/24)\,w-3 \, m^2 = 0\, .
\end{equation}
%So, the one--dimensional Ramond representation at fixed
%$c$ and $\cpl$ is labeled
%by two weights, e.g. by $h$ and $w$.
Taking $h\!=\!c/24$ one gets $G_0\!=\!0$. Such 
representation is called Ramond ground state.
In the limit $h \to c/24$ the weight $m$ approaches $0$ like 
$m \sim \left( 8 \, (h- c/24)  \, \www \, w /3  \right)^{1/2}$
and then $M_0=0$ and $H_0=\sqrt{2 \, \www \, w /3}\,$.

The definition of
the two--dimensional representations of the zero mode algebra
(\ref{zmaGG}--\ref{zmaUU}) is more complicated.
The bosonic zero modes $L_0$, $M_0$ and $W_0$ can not
be diagonalized simultaneously, since $M_0$ and $W_0$
do not commute, even ``effectively''. 
One can label the \hwr s by $h$, $w=\half\text{Trace}(W_0)$
and $m=\half\text{Trace}(M_0)$. In the following
sections we will use another labels, but they
will be always linearly dependent on the $h$, $w$, $m$.
The maximal set
of commuting operators contains 3 operators, which  
can be chosen as following:
$L_0$, some fermionic operator $F_0$ and its square
$F_0^2$. 
%At fixed $c$ and $\cpl$ the two--dimensional
%representation is labeled by 3 independent parameters:
%The $L_0$ eigenvalue $h$, one of the two eigenvalues
%of $F_0^2$ and one more parameter, which has no obvious
%meaning in the general case and will be specified in the next 
%section. 
In section~\ref{G2 algebra} it will be convenient to
choose the $F_0$ operator as the zero mode of the \none
subalgebra supersymmetry generator. 
%but we also 
%postpone the exact definition to the next section.

%%%%%%%%%%%%%%%%%%%%%%%%%%%%%%%%%%%%%%%%%%%%%%%%%%%%%%%%%%%%%%%%

\subsection{First twisted sector}

%%%%%%%%%%%%%%%%%%%%%%%%%%%%%%%%%%%%%%%%%%%%%%%%%%%%%%%%%%%%%%%%

There are 3 zero modes: $L_0$, $W_0$ and $H_0$.
The zero mode
algebra is obtained 
by setting $\cpl=0$ in (\ref{zmaHH}) and (\ref{zmaHW}):
\begin{align} 
\acom{H_0}{H_0}      
&=  2\, (h - c/24) + 4/3 \, \www  \, W_0 \, ,\\ 
\com{H_0}{W_0}      
&=  0 \, .
\end{align}
%
%The zero mode algebra is given in appendix~\ref{zemoal_firsttw}.
Its irreducible representations are one dimensional.
The \hws\ is labeled by two weights:
\begin{equation}
\begin{aligned}
L_0 \, \ketTW{h}{w} &= h \ketTW{h}{w} \, ,\\
W_0 \, \ketTW{h}{w} &= w \ketTW{h}{w} \, ,\\
H_0 \, \ketTW{h}{w} &= 
\left((h-c/24)+2/3 \, \www \, w \right)^{1/2}  \ketTW{h}{w} \, .
\end{aligned}
\end{equation}

%%%%%%%%%%%%%%%%%%%%%%%%%%%%%%%%%%%%%%%%%%%%%%%%%%%%%%%%%%%%%%%%

\subsection{Second twisted sector}

%%%%%%%%%%%%%%%%%%%%%%%%%%%%%%%%%%%%%%%%%%%%%%%%%%%%%%%%%%%%%%%%

There are four zero modes: $L_0$, $W_0$, $G_0$ and $U_0$.
Again $L_0$ is commutative with other zero modes 
and represented by its eigenvalue $h$.
The commutation relations are 
%presented in 
%appendix~\ref{zemoal_secondtw}. 
\begin{align} 
\acom{G_0}{G_0}
&=  2\, (h - c/24)  \, ,\\ 
\com{G_0}{W_0}      
&= U_0 \, ,\\
\acom{G_0}{U_0}
&= 0\, , \\
\com{W_0}{U_0}      
&= \frac{1}{2\,c}
\Big( 27\, (9/48-(h-c/24))\, G_0
+9\,\www \, (U_0-2\, G_0 \, W_0) \Big) \, ,\\
\acom{U_0}{U_0}
&=\frac{1}{c}
\Big(
27\, (h-c/24)\,(9/48-(h-c/24)) \Big. \nonumber \\
& \quad \Big.
-18\, \www \,(h-c/24)\, W_0 +9\,\www\,G_0\,U_0 \Big) \, .
\end{align}
The last two commutators are ``effective''.

Similarly to the Ramond sector the irreducible
representations of the zero mode algebra are
one or two--dimensional. They are labeled by two weights.
In the one--dimensional
representation the zero modes are
\begin{equation}
W_0=w\, , \qquad G_0=(h-c/24)^{1/2} \, , \qquad 
U_0=0\, .
\end{equation}
The representation exists only if
\begin{equation}                \label{cond_1dim_tw2}
h-c/24=0  \qquad \text{or}  \qquad 
9 + 2\,c - 48\,h - 32\, \www \, w=0 \, .
\end{equation}
%So, at fixed central charge $c$ (remember: $\cpl=0$)
%the one--dimensional \hws\ is labeled by one
%independent parameter only ($h$ or $w$). 

The two--dimension representation is constructed 
in the following way. One can choose to diagonalize
$L_0$ and $W_0$, and then the zero mode algebra is
satisfied by
\begin{equation}
\barray{c}
W_0=
\left(
\barray{cc}
w_1 & 0\\
0 & w_2\\
\earray
\right),
\qquad
G_0=\left(
\barray{cc}
0&g\\
g&\, 0\\
\earray
\right) ,
\qquad
U_0=\left(
\barray{cc}
0&u\\
-u&\, 0\\
\earray
\right),\\[11pt]
g=(h-c/24)^{1/2}, \qquad u=g\,(w_2-w_1). %,\\[3pt]
\earray
\end{equation}
There is a connection between $w_1$ and $w_2$:
\begin{equation}   \label{connec w1_w2}
2\,c\,(w_1-w_2)^2-9\,\www \,(w_1+w_2)-27\,(h-c/24)+81/16=0\, .
\end{equation}
The two--dimensional \hws\ is labeled by two weights:
$h$ and $w_1$.

%%%%%%%%%%%%%%%%%%%%%%%%%%%%%%%%%%%%%%%%%%%%%%%%%%%%%%%%%%%%%%%%

\section{Minimal models}

\setcounter{equation}{0}

\label{minimal models}

%%%%%%%%%%%%%%%%%%%%%%%%%%%%%%%%%%%%%%%%%%%%%%%%%%%%%%%%%%%%%%%%

\subsection{Unitary representations}

%%%%%%%%%%%%%%%%%%%%%%%%%%%%%%%%%%%%%%%%%%%%%%%%%%%%%%%%%%%%%%%%

As we have shown in section~\ref{N=1 superconformal subalgebras}
the existence of \none superconformal subalgebras
restricts the values of $c$ and $\cpl$ corresponding to unitary 
models of the \sww algebra.

The unitary \hwr s of an algebra are unitary
with respect to all its real subalgebras.
So there are also restrictions on the 
weights of unitary \hwr s, coming
from the non-unitarity theorem (appendix~\ref{appN=1})
of the N=1 superconformal algebra.

In the region III of $c$, $\cpl$ values (\ref{c cpl c}, \ref{c cpl cpl})
there are 3 different \none \sprc subalgebras.
The NS representation is labeled by three weights:
$h$, $w$, $m$, which are linear functions of 
three weights $\dm{1}$, $\dm{2}$, $\dm{3}$ of the three
\none subalgebras of central charge $\ck{k_1}$,
$\ck{k_2}$, $\ck{k_1+k_2}$ respectively. 
The connection between the two sets of weights is taken
from (\ref{h_to_ddd}, \ref{m_to_ddd}, \ref{w_to_ddd}).
The necessary condition for NS representation 
to be unitary is that the weights $\dm{1}$, $\dm{2}$, $\dm{3}$
are included in the correspondent 
Kac tables~(\ref{eq:N=1_dimensions})
of conformal dimensions of the \none \sprc algebra.
Therefore there is a finite
number of unitary \hwr s in the region III models,
and we can call them the minimal models
of the \sww algebra.

The \hwr\ of the \sww algebra can be decomposed
to  the sum of representations of the \none
subalgebra. Let's take for example the third subalgebra
($c=\ck{k_1+k_2}$). As we have shown in  
section~\ref{N=1 superconformal subalgebras} 
the generators of the \sww algebra fall to the
$\Phi_{1,1}$ and $\Phi_{1,3}$ representations of the 
third subalgebra.
The fusion rule for $\Phi_{1,3}$ is
\begin{equation}
 \Phi_{1,3} \times  \Phi_{m,n} = 
\Phi_{m,n-2}+\Phi_{m,n}+\Phi_{m,n+2}\, ,
\end{equation}
therefore the \hws\ of the \sww algebra with
$\dm{3}=d_{m,n}$ has \desc s lying
in the $\Phi_{m,n\pm 2}$, $\Phi_{m,n\pm 4}$, \ldots
representation of the third subalgebra,
and the \sww \hwr\ is decomposed to the sum of the third 
subalgebra representations with dimensions from the same row
of the Kac table. 

Applying these conclusions to the Ramond sector
one gets that $\tsub^{(3)}_0$ in the two--dimensional
representation looks like 
$\left(
\barray{cc}
d_{m,n} & 0\\
0& d_{m,n+2}
\earray
\right)$
in the basis, where it is diagonal; $d_{m,n}$
is the dimension of Ramond representation of the \none \sprc algebra.

Although the zero modes of the three \none subalgebras
$\tsub^{(1)}_0$, $\tsub^{(2)}_0$, $\tsub^{(3)}_0$,
can not be diagonalized simultaneously, one still can  
label the Ramond representation by 3 pairs of
Ramond dimensions (nearest Ramond neighbors in the 
correspondent Kac tables). Taking trace of zero modes
in (\ref{h_to_ddd}) one obtains the conformal dimension
\begin{align}
 h&= \frac{1}{4}\,\Big( - 
   (\dm{3}_{m_3,n_3}+ \dm{3}_{m_3,n_3+2})\left({k_1} 
+{k_2} + 2 \right) 
      \Big. \nonumber \\
& \Big. \quad + (\dm{1}_{m_1,n_1}+ \dm{1}_{m_1+2,n_1})
      \left({k_1} + 4 \right)  + 
    (\dm{2}_{m_2,n_2}+ \dm{2}_{m_2+2,n_2}) \left({k_2} + 4 \right)
\Big).    \label{h ddd R}
\end{align}

For the twisted sectors the situation is different.
The twisted representations exist only in
the minimal models with $k_1=k_2=k$. Since
one cannot mix $G$ and $H$ generators in 
the twisted sector, the only \none subalgebra
is the third subalgebra, its  generators
are given by (\ref{gsub_cpl_zero}) and (\ref{tsub_cpl_zero}).
The two weights of the \hwr\ in the tw1 sector can be
chosen to be the $\tsub^{(3)}_0$ eigenvalue $\dm{3}$
and the conformal dimension $h$. The tw1 representation
is of Ramond type with respect to the third \none subalgebra.
The conditions (\ref{cond_1dim_tw2}) for existence
of the one dimensional representation in the tw2 sector
of the $k_1=k_2=k$ minimal model
are  rewritten as 
\begin{equation}
  h-c/24=0 \qquad \text{or}  \qquad \dm{3}=d_{k+2,k+2} \, .
\end{equation}
The two--dimensional representation of the tw2 type
is labeled by two weights: $h$ and a pair 
$(\dm{3}_1, \dm{3}_2)$
of the nearest NS dimensions as the eigenvalues of the
$\tsub^{(3)}_0$ operator. The connection 
(\ref{connec w1_w2}), being rewritten in terms of
$\dm{3}_1$ and $\dm{3}_2$, states
exactly that  $\dm{3}_1$ and $\dm{3}_2$ are nearest NS neighbors
in the row of the correspondent Kac table.

All the conditions of unitarity so discussed are not 
sufficient. In the case $k_1=1$ one gets additional
restrictions on unitary representations by
noting that the energy momentum tensor
can be decomposed to two commuting parts
\begin{equation}
  T=\tsub^{(2)} + T^{\text{Vir}}_{k_2+2}
   \qquad  \text{or} \qquad 
  T=\tsub^{(3)} + T^{\text{Vir}}_{k_2}\, ,
\end{equation}
where $T^{\text{Vir}}_{k}$ is the generator of the Virasoro
algebra of central charge 
\begin{equation}
c=c_k^{N\! = \! 0}=1-\frac{6}{(k+2)(k+3)}\, , 
\end{equation}
corresponding to
the unitary minimal models of the Virasoro algebra 
\cite{BPZ}. (One could see the decomposition from (\ref{decom3}) and
(\ref{decom1}).)
This fact restricts the values of $h-\dm{2}$
and $h-\dm{3}$. For the $k_1=1$ models all
the discussed restrictions on the
weights are in fact sufficient conditions of unitarity
of NS and Ramond representations. (We do not prove it here.)
The discussion for $k_1=1$ applies to the $k_2=1$
case as well.

By taking formally $k_1=0$ one should obtain
the minimal models of \none \sprc algebra
(appendix~\ref{appN=1}).

In addition we know the examples (see section \ref{Examples})
 of explicit construction of $k_1=k_2=1$ and $k_1=k_2=2$
minimal models of the \sww algebra in terms of 
$N\!=\!2$ models.

Based on these facts and with the help of the
coset construction (\ref{coset})
we guess the unitary spectrum of the general
(arbitrary $k_1$ and $k_2$)
\sww minimal model.
The full list of minimal model
unitary representations is presented in 
appendix~\ref{list rep}.
The unitarity was also checked
by computer calculations of the \sww algebra
Kac determinant on the few first levels.

The list of NS and Ramond representations
forms a three--dimensional table with indices
$s_1,s_2,s_3$, running in the range (\ref{s range}).
The twisted sector representations form a two--dimensional
table with indices $t_1$ and $t_2$ (\ref{t range}).
There is a same number of NS and Ramond representations
and the same number of tw1 and tw2 representations.

Substantial part of the spectrum could be predicted
using the magic relation between the dimensions
of any \none minimal model (\ref{eq:N=1_dimensions}):
\begin{equation}
d^k_{m,1}+d^k_{1,n}-d^k_{m,n}=\frac{(m-1)(n-1)}{4}
\qquad \forall k.
\end{equation}
The relation is to be understood in the context
of the fusion rule
\begin{equation}
\Phi_{m,1} \times \Phi_{1,n}=\Phi_{m,n} \, .
\end{equation}
Taking $m=3$ we get, that $\Phi_{1,n}$
fields are local or semilocal with respect
to $\Phi_{3,1}\,$. 
The \sww algebra is decomposed to 
$\Phi_{1,1} \oplus \Phi_{3,1}$ representation
of the first \none subalgebra. Therefore
the field $\Phi_{1,n}$ (of the first \none subalgebra)
is a valid representation of the whole \sww algebra, since
it is local (or semilocal) with respect to all the generators
of \sww. This representation is of Ramond or NS type,
depending on $n$ is even or odd respectively.
The conformal dimension $h$ of such a field coincides
with the weight of the first \none subalgebra.
For $n \le 4$ the $\Phi_{1,n}$ field lies in the
\hwr\ of the \sww algebra, for $n>4$ it is \desc\
of some \hwr . We call such a representation
the purely internal representation with respect
to the first \none subalgebra. Obviously the set
of purely internal representations is closed under fusion rules.
(The similar situation is encountered in the case of the \sw
algebra, which have purely internal representations
with respect to its Virasoro subalgebra (section 4 of \cite{gn}).)
Of course, there are also purely internal representations
with respect to the second and to the third \none subalgebras.
The representations $(s_1,1,1)$, $(1,s_2,1)$, $(1,1,s_3)$
are purely internal of the first, second and third subalgebras
respectively. Such representations have simple fusion rules.

It is interesting to note, that $(1,s_2,s_3)$
fields of the \sww algebra are
local or semilocal with respect to 
the $(3,1,1)$ field.

%%%%%%%%%%%%%%%%%%%%%%%%%%%%%%%%%%%%%%%%%%%%%%%%%%%%%%%%%%%%%%%%

\subsection{Fusion rules}

\label{Fusion rules}

%\setcounter{equation}{0}

%%%%%%%%%%%%%%%%%%%%%%%%%%%%%%%%%%%%%%%%%%%%%%%%%%%%%%%%%%%%%%%%

It is well known that the fusion rules of NS and
Ramond sector representations have $\mathbb{Z}_2$
grading: (NS $\to 0$, Ramond $\to 1$ under addition modulo 2).
The \sww algebra has two additional twisted sectors:
tw1 and tw2. The full set of fusion rules has
$\mathbb{Z}_2 \times  \mathbb{Z}_2$ grading:
NS $\to (0,0)$, Ramond $\to (1,0)$, tw1 $\to (0,1)$,
tw2 $\to (1,1)$.
The fusion rules of different sectors are  summarized in the table:
\begin{center}
\begin{tabular}{|c|cccc|}
 \hline
   & NS & R & tw1 & tw2 \\
 \hline
NS & NS & R & tw1 & tw2 \\
% \hline
R  & R  & NS& tw2 & tw1 \\
% \hline
tw1&tw1 &tw2& NS  & R   \\
% \hline
tw2&tw2 &tw1& R   & NS  \\
 \hline
\end{tabular}
\end{center}

The fusions of the \sww representations 
have to be consistent with fusion rules
of its subalgebras. In the case of minimal models
there are three \none \sprc subalgebras and
the fusions of the \sww NS and Ramond representations
are completely fixed by the fusions of the correspondent
\none minimal models (\ref{eq:N=1_fusions}).
The fusion of $({s'_1,s'_2,s'_3})$ and $({s''_1,s''_2,s''_3})$ 
representation (see appendix~\ref{list rep})
of the \sww algebra is
\begin{equation}             \label{eq:sww_fusions}
({s'_1,s'_2,s'_3}) \times ({s''_1,s''_2,s''_3}) = 
\sum_{s_1=|s'_1-s''_1|+1}^
{\min(s'_1+s''_1-1, \atop
2k_1+3-(s'_1+s''_1))}
%{s'_1+s''_1-1} 
\,\,
\sum_{s_2=|s'_2-s''_2|+1}^
{\min(s'_2+s''_2-1, \atop
2k_2+3-(s'_2+s''_2))}
%{s'_2+s''_2-1} 
\,\,
\sum_{s_3=|s'_3-s''_3|+1}^
{\min(s'_3+s''_3-1, \atop
2(k_1+k_2)+7-(s'_3+s''_3))}
%{s'_3+s''_3-1} 
%\,\,
\!\!\!\!\!\!\!\!
({s_1,s_2,s_3}) \, ,
\end{equation}
where $s_1$, $s_2$ and $s_3$ are raised by steps of 2.
The selection of one index is independent on two others
and satisfies the ``$\sutwo$ pattern''. The fusion
rules of the $N\!=\!0$ and the \none minimal models
satisfy the same pattern, the only difference
is that in our case the table of representations
is three--dimensional. 

The proof is based on the first column of formula 
(\ref{m n min_mod}). The $({s_1,s_2,s_3})$
representation is decomposed to the sum of 
 representations of the first \none subalgebra
from the column number $s_1$ of the correspondent \none Kac table.
Thus the column selection rule of the first \none
subalgebra is preserved and coincides with the $s_1$
selection in (\ref{eq:sww_fusions}). Similarly
the $s_2$ and $s_3$ selection rules are adopted 
from the column and row selection rules of
the second and the third \none subalgebras respectively.

The twisted representations of the minimal models
are labeled by two numbers: $t_1$ and $t_2$ 
(appendix~\ref{list rep}). But there is only one
\none subalgebra (the third one) in the twisted sector.
One can reed from ({\ref{list tw reps}}) 
that $t_1$ is the row number in the correspondent Kac table,
meaning that $t_1$ has common selection rules with $s_3$.
The selection of $t_2$ in the fusion rules can not be 
fixed by the described methods.

The ``corner'' entries of the three--dimensional
table of NS and Ramond representations are
the $(1,1,1)$, $(k_1+1,1,1)$, $(1,k_2+1,1)$, 
$(1,1,k_1+k_2+3)$ representations.
The first one is the vacuum representation.
The three others have the following fusion ``square'':
\begin{equation}
  \Phi \times \Phi =I,
\end{equation}
where $I$ denotes the identity (vacuum) representation.
If such a field $\Phi$ is of the Ramond type, then
the fusion of it with the other fields defines
one-to-one transformation,
mapping NS fields to Ramond ones. (This is 
an analogy of the $U(1)$ flow of the
$N\!=\!2$ superconformal algebra \cite{Schwimmer:mf}.) 
If the field $\Phi$ is of the NS type, then 
its conformal dimension $h$ is integer or half--integer and
the \sww algebra can be extended to include this field.
In the case, then both $k_1$ and $k_2$ are even,
all three ``corner'' fields are of the NS type;
in other cases (at least one $k$ is odd) 
one ``corner''
fields is of the NS type and
two are of the Ramond type (and
then there are two different NS--R isomorphisms).

%%%%%%%%%%%%%%%%%%%%%%%%%%%%%%%%%%%%%%%%%%%%%%%%%%%%%%%%%%%%%%%%

\subsection{Examples}

\label{Examples}

%\setcounter{equation}{0}

%%%%%%%%%%%%%%%%%%%%%%%%%%%%%%%%%%%%%%%%%%%%%%%%%%%%%%%%%%%%%%%%

The following two examples are the $c=3/2$, $\cpl=0$
($k_1=k_2=1$) and  $c=9/4$, $\cpl=0$ ($k_1=k_2=2$)
minimal models of the \sww algebra. The former model is realized
as $Z_2$ orbifold of a tensor product of free fermion and free 
boson on radius $\sqrt{3}$, the latter is realized
as $Z_2$ orbifold of the sixth ($c=9/4$) minimal model
of the $N=2$ superconformal algebra. 
% Before we go to the examples we introduce the
% $N=2$ superconformal algebra. The algebra is
% generated by energy--momentum operator $T$,
% two supersymmetry generators $G^+$ and $G^-$
% of spin $3/2$,
% and $U(1)$ current $J$ of spin 1.
% The \ope s are:
% \begin{align}
% J(z) \, J(w) &= \frac{c/3}{(z-w)^2} \, ,\\
% J(z) \, G^{\pm}(w)&=\frac{\pm G^{\pm}(w)}{z-w} \, ,\\
% G^+(z) \, G^-(w)&=\frac{2\,c/3}{(z-w)^3}+
% \frac{2\,J(w)}{(z-w)^2}+\frac{2\,T(w)+\dd J(w)}{z-w} \, ,\\
% G^+(z) \, G^+(w)&=G^-(z) \, G^-(w)=0 \, .
% \end{align}
   
%%%%%%%%%%%%%%%%%%%%%%%%%%%%%%%%%%%%%%%%%%%%%%%%%%%%%%%%%%%%%%

\subsubsection{$c=3/2$, $\cpl=0$ model}

\label{k1_k2_1}

%%%%%%%%%%%%%%%%%%%%%%%%%%%%%%%%%%%%%%%%%%%%%%%%%%%%%%%%%%%%%%

The boson on radius $\sqrt{3}$
is equivalent to the first minimal model 
of the $N=2$ superconformal algebra.
The generators of \sww are constructed in 
the following way:
\begin{equation}                 \label{c32}
\begin{aligned}
T&=T^{N=2}+T^{\mathrm{Ising}} \, , \\
G&=\sqrt{3}\, J \, \psi \, , \\
H&=\sqrt{3/2} \, G_1 \, , \\
\end{aligned}
\qquad
\begin{aligned}
M&=-\ii \, 3/\sqrt{2} \, G_2 \, \psi \, , \\
W&=1/\sqrt{2}\, T^{N=2}-\sqrt{2}\,T^{\mathrm{Ising}} \, , \\
U&=\sqrt{2/3} \left( 2\,J\,\dd \psi-\dd J\, \psi \right) , \\
\end{aligned}
\end{equation}
where $T^{N=2}$, $G_1$, $G_2$ and $J$
are the (real) generators of the $N=2$ superconformal algebra
at $c=1$, $\psi$ is the free fermion field and $T^{\mathrm{Ising}}$
is its energy--momentum operator. The expressions 
in (\ref{c32})
are invariant under $\mathbb{Z}_2$ transformation
\mbox{$J \to -J, \psi \to -\psi$}.
One can build all the \hwr s in the model
(5 in NS, 5 in Ramond, 3 in every twisted sector)
%-- too long to be reproduced here explicitly)
by appropriate combinations of 
representations of the $N=2$ minimal model and the Ising model.
The list of representations is presented in 
table~\ref{table_k_1=k_2=1} in appendix~\ref{list rep}.

%%%%%%%%%%%%%%%%%%%%%%%%%%%%%%%%%%%%%%%%%%%%%%%%%%%%%%%%%%%%%%

\subsubsection{$c=9/4$, $\cpl=0$ model}

%%%%%%%%%%%%%%%%%%%%%%%%%%%%%%%%%%%%%%%%%%%%%%%%%%%%%%%%%%%%%%

The NS sector of the sixth ($c=9/4$)  $N=2$ minimal model
contains \hws\ $\Psi_6^0$ of conformal dimension $3/2$ and
zero $U(1)$ charge. The $N=2$ superconformal algebra
can be extended by $N=2$ superprimary
field corresponding to this state \cite{Inami:1989yi}.
Then the fields, invariant under $\mathbb{Z}_2$
transformation \mbox{$J \to -J$}
form the \sww algebra:
\begin{equation}      \label{c94}
\begin{aligned}
T&= T^{N=2} \, , \\
G&= G_1 \, , \\
\end{aligned}
\qquad
\begin{aligned}
H&=\Psi_6^0 \, , \\
M&=\Phi_6^0  \, , \\
\end{aligned}
\qquad
\begin{aligned}
W&=(2 \, T- 3 \, \NO{J}{J})/\sqrt{5} \, , \\
U&=2 /\sqrt{5} \, (3\, \ii \, \NO{J}{G_2}-\dd G_1) \, ,
\end{aligned}
\end{equation}
where again $T^{N=2}$, $G_1$, $G_2$ and $J$
are the (real) generators of the $N=2$ superconformal algebra
and $\ket{\Phi_6^0}$ is a superpartner of $\ket{\Psi_6^0}$: 
$\ket{\Phi_6^0}=({G_1})_{-\half} \ket{\Psi_6^0}$.
The \hwr s of the  $N=2$ 
minimal model can be easily transformed to representations
of the \sww algebra.
The list of \hwr s consists of 16 NS, 16 Ramond, 6 tw1
and 6 tw2 representations (too long to be reproduced here explicitly).

%%%%%%%%%%%%%%%%%%%%%%%%%%%%%%%%%%%%%%%%%%%%%%%%%%%%%%%%%%%%%%%%

\section{Spectrum of the $G_2$ conformal algebra}

\label{G2 algebra}

\setcounter{equation}{0}

%%%%%%%%%%%%%%%%%%%%%%%%%%%%%%%%%%%%%%%%%%%%%%%%%%%%%%%%%%%%%%%%

In this section we discuss the \sww algebra
at $c=10\half, \cpl=0$. As we have shown in 
section~\ref{N=1 superconformal subalgebras}
there is one real \none subalgebra. It has central charge
$\csub=7/10$ and thus coincides with the
tricritical Ising model. The subalgebra is
generated by operators $\gsub$ (\ref{gsub_cpl_zero}) 
and $\tsub$ (\ref{tsub_cpl_zero}).
The \sww algebra is decomposed 
to the $\Phi_{1,1} \oplus \Phi_{3,1}$
representation of the \none subalgebra.
Since at $\csub=7/10$ the fields
$\Phi_{3,1}=\Phi_{2,2}$ are identical
there is a new null state on level $2$: 
\begin{equation}       \label{null state}
3\,\gsub_{-3/2}\,\gsub_{-1/2}-2\,\tsub_{-2}\,\ket{\Phi_{3,1}} .
\end{equation}
(The null state on level $3/2$ is already encoded in the
structure of the \sww algebra, but the existence
of the null state (\ref{null state}) is a special
feature of the tricritical Ising model.)
Translating (\ref{null state}) to the language of generators of
the \sww algebra one gets that the null field is
\begin{equation}
  \label{eq:ideal}
  2\,{\sqrt{14}}\,\NO{G}{W} - 
  3\,\NO{H}{M} + 
  2\,\NO{T}{G} - 
  2\,{\sqrt{14}}\,\dd U  .
\end{equation}
This is an ideal of the \sww algebra at  $c=10\half, \cpl=0$.
(The existence of the ideal is known 
since \cite{Figueroa-O'Farrill:1996hm}.)

In ref.~\cite{Shatashvili:zw} the conformal algebra associated
to the manifolds of $G_2$ holonomy is derived.
Up to the ideal (\ref{eq:ideal}) the $G_2$ conformal algebra
coincides with the \sww algebra at $c=10\half, \cpl=0$.
In the free field representation, used by the authors of
\cite{Shatashvili:zw} to obtain the $G_2$ algebra,
the ideal (\ref{eq:ideal}) vanishes identically.
The authors of ref.~\cite{Shatashvili:zw} used
different basis
of generators of the algebra. Their basis
is connected to ours by
\begin{equation}
\begin{aligned}
\Phi &=\ii \, H ,\\
K &= \ii \, M ,
\end{aligned}
\qquad
\begin{aligned}
X &=-(T+\sqrt{14}\,W)/3 ,\\
\tilde M &= -(\dd G+2\,\sqrt{14}\,U)/6 .
\end{aligned}
\end{equation}
The $T$ and $G$ generators are the same.
We have explicitly checked, that the \ope s
in the first appendix of \cite{Shatashvili:zw}
coincide (up to the ideal (\ref{eq:ideal}))
with the \ope s of the \sww algebra.

Some unitary \hwr s of the $G_2$ algebra are found in
\cite{Shatashvili:zw}. In this section we complete the list
of unitary representations. Our calculation is based
on the fact, that the $T, G, W, U$ fields
of the $G_2$ algebra generate closed subalgebra
modulo the same ideal (\ref{eq:ideal}) and its 
\desc s. This subalgebra is the \sw superconformal algebra
\cite{fofs,
Komata:1991cb,
Blumenhagen:1992nm,
Blumenhagen:1992vr,
Eholzer:1992pv,
Mallwitz:1994hh,
gn}
of central charge $10\half$. The $G_2$ algebra can be seen
as an extended version of the \sw algebra.
It is interesting to note that the \sw algebra at 
another value of central charge
($c=12$) is the superconformal algebra associated
to manifolds of $Spin(7)$ holonomy \cite{Shatashvili:zw}.

The complete spectrum of unitary representations
of the \sw algebra is found in \cite{gn}. 
The $c=10\half$ unitary model spectrum is presented
in table~\ref{sw spectrum}, where
$x$ stands for real positive number in the 
continuous spectrum representations.
\begin{table}[tp]
%\begin{equation}
$$
\begin{array}{|c|c||c|c|}
\hline
\multicolumn{2}{|c||}{\text{NS}}
&
\multicolumn{2}{|c|}{\text{Ramond}}\\
\hline
h&a&h&a\\
\hline 
0 &0 & 7/16 & 0\\
3/8 & 3/80 & 7/16 & 3/80\\
1/2 & 1/10 & 7/16 & 1/10 \\
7/8 & 7/16 & 7/16 & 7/16 \\
1 &3/5 &15/16 & (1/10,3/5)\\
3/2 & 3/2& 31/16 & (3/5,3/2)\\
\hline
x&0 &7/16+x& (0,1/10)\\
3/8+x & 3/80 &7/16+x&  3/80\\
1/2+x & 1/10 &7/16+x& (3/80,7/16)\\
1+x &3/5 &15/16+x&(1/10,3/5)\\
\hline
\end{array}
$$
%\end{equation}
\caption{The $c=10\half$ unitary model 
of the \sw algebra.}  \label{sw spectrum}
\end{table}
There are NS and Ramond representations,
they are labeled by two numbers:
the conformal dimension $h$ and the
internal dimension $a$ (the weight of 
the $c=7/10$ Virasoro subalgebra of the \sw algebra).

The $H$ field 
of the \sww algebra
is identified with
the $h=3/2,a=3/2$ NS representation of the \sw algebra,
$M$ is its superpartner.

The $(H,M)$ supermultiplet is purely internal
with respect to the $c=7/10$ Virasoro subalgebra
of the \sw algebra. Due to this fact we know its 
fusion rules with all other representations
of the \sw algebra. In order to get the 
representations of \sww one have to combine 
the table~\ref{sw spectrum} representations 
in multiplets under the action of the
$h=3/2, a=3/2$ field. It can be easily done,
however this way one gets only two weights. But we
know that the \sww NS and Ramond representations 
are labeled by three weights. It is convenient
to choose them: the conformal dimension $h$;
the weight of the \none subalgebra, which
coincides with the \sw algebra internal dimension $a$;
and the eigenvalue $m$ of the $M$ zero mode.
(In the case of two--dimensional Ramond
representation $m$ stands for $\half \text{Trace}(M_0)$.)
The twisted representations are labeled by two weights:
$h$ and $a$. 

One obtains the third weight $m$ by using
again the null field (\ref{eq:ideal}).
Acting by it on the \hws\ one should get zero.
Consider, for example, the NS sector.
Take the $G_{-1/2}$ \desc\ of the ideal:
\begin{equation}
  2{\sqrt{14}} \left(
2\NO{T}{W}-\NO{G}{U}-\dd^2 W \right)
- 
  \NO{G}{\dd G} + 
  3\NO{H}{\dd H} - 
  3\NO{M}{M} + 
  4\NO{T}{T} + \dd^2 T  .
\end{equation}
The eigenvalue of its zero mode
should be set to zero for a 
consistent representation.
This leads to a connection between
$m$ and two other weights $h$ and $a$:
\begin{equation}
  m^2=10\,a\, (2\,h-1) .
\end{equation}
The similar connection can be found in the Ramond case.
Concluding, the unitary representations of the
$G_2$ algebra are (again $x>0$):\\
NS sector:
\begin{equation}
  \label{eq:g2 spectrum NS}
\begin{array}{ll}
1) & h=a=m=0 , \\
2) & h=1/2, \quad  a=1/10 ,\quad  m=0 , \\
3) & h=x ,\quad  a=0 ,\quad  m=0 , \\
4) & h=1/2+x ,\quad  a=1/10 ,\quad  m=\sqrt{x} . 
\end{array}
\end{equation}
Ramond sector:
\begin{equation}
  \label{eq:g2 spectrum R}
\begin{array}{ll}
1) & h=7/16, \quad a=7/16, \quad m=0 , \\
2) & h=7/16, \quad a=3/80, \quad m=0 , \\
3) & h=7/16+x, \quad a=(3/80,7/16), \quad m=0 , \\
4) & h=7/16+x, \quad a=3/80, \quad m=\sqrt{x}/2 .
\end{array}
\end{equation}
tw1 sector:
\begin{equation}
  \label{eq:g2 spectrum tw1}
\begin{array}{ll}
1) & h=3/8, \quad a=3/80 , \\
2) & h=7/8, \quad a=7/16 , \\
3) & h=3/8+x, \quad a=3/80 . 
\end{array}
\end{equation}
tw2 sector:
\begin{equation}
  \label{eq:g2 spectrum tw2}
\begin{array}{ll}
1) & h=7/16, \quad a=1/10 , \\
2) & h=7/16, \quad  a=0 , \\
3) & h=7/16+x ,\quad  a=(0,1/10) . 
\end{array}
\end{equation}
The NS  and Ramond discrete spectrum states
(the first two in (\ref{eq:g2 spectrum NS})
and (\ref{eq:g2 spectrum R})) were found in
\cite{Shatashvili:zw}.

The first Ramond representation 
($h=7/16, a=7/16, m=0$) is purely
internal with respect to the tricritical
Ising model. Due to this fact we know its fusions.
Since its square is identity, the field serves
as an isomorphism mapping, connecting NS and Ramond
sectors, and  connecting the tw1 and tw2 sectors.
The representations of the same line
number in (\ref{eq:g2 spectrum NS}) and
(\ref{eq:g2 spectrum R}) and in 
(\ref{eq:g2 spectrum tw1}) and (\ref{eq:g2 spectrum tw2})
are isomorphic.

%%%%%%%%%%%%%%%%%%%%%%%%%%%%%%%%%%%%%%%%%%%%%%%%%%%%%%%%%%%%%%%%

\section{Summary}

%%%%%%%%%%%%%%%%%%%%%%%%%%%%%%%%%%%%%%%%%%%%%%%%%%%%%%%%%%%%%%%%

In this paper we study the \sww superconformal algebra.
%The connection of this algebra with the coset space (\ref{coset})
%is proved by explicit construction.
We show by explicit construction
that the coset (\ref{coset})
contains the \sww algebra.
%This coset represents a wide set of rational
%conformal field theories possessing the \sww
%algebra symmetry.
The space of parameters ($c$ and $\cpl$) is
divided to two regions (figure~\ref{figure1}).

In the first region there are unitary models at
discrete points in the $c, \cpl$ space.
These are the minimal models of the
algebra, they are described by the 
coset (\ref{coset}).
In this region the \sww algebra has
three different nontrivial \none subalgebras.
%The weights of \hwr s are chosen to be
The conformal dimensions with respect
to these subalgebras serve as the weights
of the \sww \hwr s. The fusion rules are also dictated
by the fusions of the \none subalgebras.
The characters 
of \hwr s are not discussed in this paper. We
suppose that the characters can be easily
obtained from the coset construction.

In the second region of parameters
the \sww algebra has one \none \sprc subalgebra.
The unitary models ``lie'' on unitary curves
and have continuous spectrum of unitary representations.
One of the continuous unitary models is the $c=10\half$,
$\cpl=0$ model. The \sww algebra at these values
of parameters coincides (up to a null field)
with the superconformal algebra, associated to
the manifolds of $G_2$ holonomy. From the other
point of view it is an extended version of the 
\sw algebra, which at another value of the central charge
($c=12$) corresponds to the manifolds of $Spin(7)$
holonomy. We find the unitary spectrum of the 
$G_2$ holonomy algebra.
The connection of various realizations
of the $G_2$ superconformal algebra
with the geometric properties
of the $G_2$ manifolds is the open problem for study.

%%%%%%%%%%%%%%%%%%%%%%%%%%%%%%%%%%%%%%%%%%%%%%%%%%%%%%%%%%%%%%%%

\subsection*{Acknowledgment}

It is pleasure to thank Doron Gepner and
Alexander Zamolodchikov
for useful discussions.

%%%%%%%%%%%%%%%%%%%%%%%%%%%%%%%%%%%%%%%%%%%%%%%%%%%%%%%%%%%%%%%%

\newpage

%%%%%%%%%%%%%%%%%%%%%%%%%%%%%%%%%%%%%%%%%%%%%%%%%%%%%%%%%%%%%%%%

%\appendix{Appendix}

%\enlargethispage*{4pt}

\renewcommand{\thesection}{Appendix}

\section{}

\renewcommand{\thesubsection}{\Alph{subsection}}

\renewcommand{\theequation}{\thesubsection.\arabic{equation}}

%%%%%%%%%%%%%%%%%%%%%%%%%%%%%%%%%%%%%%%%%%%%%%%%%%%%%%%%%%%%%%%%

\subsection{%Operator product expansions 
\ope s of the \sww algebra}
\label{apA}

\setcounter{equation}{0}

%%%%%%%%%%%%%%%%%%%%%%%%%%%%%%%%%%%%%%%%%%%%%%%%%%%%%%%%%%%%%%%%

$T$ and $G$ generate \none superconformal 
algebra of central charge $c$.
$(H,M)$ and $(W,U)$ are its superprimary fields.
The nontrivial \ope s are:
{\small
\begin{align}
H(z)\, H(w)&=
\frac{\frac{2\, c}{3}}{(z-w)^3}+
\frac{\cpl\, M + 2\, T + \frac{4\, {\www}}{3}\, W}{z-w} \, , \\
H(z)\, M(w)&=
\frac{3\, G + 3\, {\cpl}\, H}{(z-w)^2}+
\frac{-\frac{2\, {\www}}{3}\, U + \dd G + 
{\cpl}\, \dd H}{z-w} \, , \\
M(z)\, M(w)&=
\frac{2\, c}{(z-w)^4}+
\frac{4\, \cpl \, M + 
8\, T + \frac{4\, {\www}}{3}\, W}{(z-w)^2}+
\frac{2\, \cpl \, \dd M + 4\, \dd T 
+ \frac{2\, {\www}}{3}\, \dd W}{z-w} \, , \\
H(z)\, W(w)&=
\frac{\www \, H}{(z-w)^2}+
\frac{\frac{\cpl }{2}\, U + 
\frac{{\www}}{3}\, \dd H}{z-w} \, , \\
\label{opeMW}
M(z)\, W(w)&=
\frac{\frac{\www}{3} \, M+ 2\, \cpl \, W}{(z-w)^2}+
\frac{\frac{9\, \www }{2\, 
      c}\, \NO{G}{H} + \frac{\www \, \left( -27 + 2\, c \right) }{12\, 
      c}\, \dd M + \cpl \, \dd W}{z-w} \, , \\
H(z)\, U(w)&=
\frac{\frac{-2\,\www }{3}\,M + 2\,\cpl \,W}{{\left( z - w \right) 
}^2} + \frac{\frac{9\,\www }{2\,c}\,\NO{G}{H} - \frac{\www 
\,\left( 27 + 2\,c \right) }{12\,c}\,\dd M + \frac{\cpl}{2}\,\dd
W}{z - w}  \, , \\
M(z)\, U(w)&=
\frac{2\,\www \,H}{{\left( z - w \right) }^3} + 
\frac{\frac{5\,\cpl }{2}\,U + \frac{2\,\www}{3} \,\dd H}{{\left( z 
- w \right) }^2}  \nonumber\\
&\hspace{15ex} +\frac{-\frac{9\,\www }{2\,c}\,\NO{G}{M} + \frac{9
\,\www }{c}\,\NO{T}{H} + \cpl \,\dd U + \frac{\www \,\left( -27 + 
2\,c \right) }{12\,c}\,\dd^2 H}{z - w}  \, , \\
W(z)\, W(w)&=
\frac{c/2}{(z-w)^4} +
\frac{2\, T+\frac{\cpl}{2}\, M 
+  \frac{\www (10\, c -27) }{6\,c}\, W}{(z-w)^2}+
\frac{\dd T+\frac{\cpl}{4}\, \dd M 
+  \frac{\www (10\, c -27) }{12\,c}\, \dd W}{z-w}  , \\
W(z)\, U(w)&=
\frac{-3\,G - \frac{3\,\cpl }{2}\,H}{{\left( z - w \right) }^3} + 
\frac{\frac{\www \,\left( -27 + 10\,c \right) }{12\,c}\,U - \dd G - 
\frac{\cpl }{2}\,\dd H}{{\left( z - w \right) }^2}  \nonumber\\
&\hspace{-4ex} -\frac{1}{48\, c\, (z - w)} 
\Big(
162\,\cpl \,\NO{G}{M} + 432\,\www \,\NO{G}{W} - 324\,\NO{H}{M} + 
648\,\NO{T}{G} + 324\,\cpl \,\NO{T}{H} \Big. \nonumber \\
& \hspace{-4ex}\Big.  - 8\,\www \,\left( 27 + 
2\,c \right) \,\dd U + 6\,\left( -27 + 2\,c \right) \,\dd^2 G + 
3\,\left( -27 + 2\,c \right) \,\cpl \,\dd^2 H \Big) \, , \\
U(z)\, U(w)&=
-\frac{2\, c}{(z-w)^5}-
\frac{\frac{5\,\cpl }{2}\,M + 10\,T + 
\frac{\www \,\left( -27 + 10\,c \right) }{3\,c}\,W}
{(z-w)^3}\nonumber \\
& \hspace{-4ex} -
\frac{\frac{5\,\cpl }{4}\,\dd M + 5\,\dd T + \frac{\www \,\left( -27 + 
10\,c \right) }{6\,c}\,\dd W}{(z-w)^2}
-\frac{1}{16\, c\, (z-w)} \Big(
-144\,\www \,\NO{G}{U} - 108\,\NO{G}{\dd G}
\Big. \nonumber \\
& \hspace{-4ex}   - 54\,\cpl \,\NO{G}{\dd H} + 
108\,\NO{H}{\dd H} - 108\,\NO{M}{M} + 216\,
\cpl \,\NO{T}{M} + 432\,\NO{T}{T} + 
288\,\www \,\NO{T}{W} 
\nonumber \\
& \hspace{-4ex} \Big.  + 54\,\cpl 
\,\NO{\dd G}{H} - 3\,\left( 9 - 2\,c \right) \,\cpl 
\,\dd^2 M + 24\,c\,\dd^2 T - 4\,\www \,
\left( 27 - 2\,c \right) \,\dd^2 W  
\Big) \, .
\end{align}
}
where 
\begin{equation}
 \www = \sqrt{\frac{9\, c\, (4+\cpl^2)}{2 \, (27-2\, c)}}  
\end{equation}
and the fields in the right hand sides of the \ope s
are taken in the point $w$.

%%%%%%%%%%%%%%%%%%%%%%%%%%%%%%%%%%%%%%%%%%%%%%%%%%%%%%%%%%%%%%%%

\subsection{Unitary minimal models of the \none superconformal 
\mbox{algebra}}
\label{appN=1}

\setcounter{equation}{0}

%\addtocounter{section}{1}

%%%%%%%%%%%%%%%%%%%%%%%%%%%%%%%%%%%%%%%%%%%%%%%%%%%%%%%%%%%%%%%%

At $c<3/2$ all unitary representations of the
\none superconformal algebra are described
by its minimal models. Their central charge is
\begin{equation}    \label{C_k}
\ck{k} = \frac{3}{2}-\frac{12}{(k+2)(k+4)}\, , 
\qquad k=0,1,2,\ldots
\end{equation} 
The conformal dimensions of the unitary \hwr s 
$\Phi_{m,n}$
of the $c=\ck{k}$ minimal model are given in
the Kac table
\begin{equation}
\begin{aligned}
  \label{eq:N=1_dimensions}
  d^{k}_{m,n}&=\frac{{\left( \left( k + 2 \right) 
\,m - \left( k + 4\right) 
\,n \right)}^2 - 4}
   {8\,\left( k + 2 \right) \,\left( k + 4 \right) }+\frac{r}{16}\, ,
\qquad
\barray{rcl}
m&=&1,2,\ldots, k+3\, , \\
n&=&1,2,\ldots, k+1\, ,
\earray
\\
r&={(m+n) \bmod 2}=\left\{
\barray{l}
0, \quad \mathrm{NS\ sector,}\\
1, \quad \mathrm{Ramond\ sector.}
\earray
\right.
\end{aligned}
\end{equation}
The fusion rules are given by $\sutwo$ like
selection rules for every index ($m$ and $n$):
\begin{equation}             \label{eq:N=1_fusions}
\Phi_{m_1,n_1} \times \Phi_{m_2,n_2} = 
\sum_{m=|m_1-m_2|+1}^
{\min(m_1+m_2-1, \atop
2k+7-(m_1+m_2))} \,\,
\sum_{n=|n_1-n_2|+1}^
{\min(n_1+n_2-1,\atop
2k+3-(n_1+n_2)) } \,\,
\Phi_{m,n} \, ,
\end{equation}
where the indices $m$ and $n$ in the sums are raised by steps of 2.

The \none minimal models correspond to the
diagonal coset construction \cite{Goddard:1986ee}:
\begin{equation}                   \label{coset N=1}
\frac{\sutwo_k \oplus \sutwo_2}{\sutwo_{k+2}} \, .
%\qquad k \in \mathbb{N}
\end{equation}

%%%%%%%%%%%%%%%%%%%%%%%%%%%%%%%%%%%%%%%%%%%%%%%%%%%%%%%

\subsection{Coset construction of the \sww algebra}
\label{apB}

\setcounter{equation}{0}

%%%%%%%%%%%%%%%%%%%%%%%%%%%%%%%%%%%%%%%%%%%%%%%%%%%%%%%%%%%%%%%%

%% k_1 \, k_2 (k_1+2) (k_2+2) (k_1+k_2+4) (k_1+k_2+6) (k_1-k_2)
We present the explicit construction of the
\sww generators in terms of the coset (\ref{coset})
currents.
For details and notations see section~\ref{sec:expli_const}.
{\small
\begin{align}
T &=
\frac{1}{ k_1+k_2+4}
\Bigg(
-\half \, (k_1+k_2) \, \NO{\psi_i}{\dd \psi_i} 
- 2 \, \NO{J_i^{(1)}}{J_i^{(2)}} \Bigg. \nonumber \\
&\Bigg. +
\bigg( \big( \, \frac{k_2+2}{k_1+2} \, \NO{J_i^{(1)}}{J_i^{(1)}} 
- 2 \,  \NO{J_i^{(1)}}{J_i^{(3)}} \big)+
\big(1 \leftrightarrow 2 \big)  \bigg)
\Bigg)
 , \\
G &=
\frac{\sqrt{2}}{\left(
(k_1+2) (k_2+2) (k_1+k_2+4)\right)^{1/2}} \times \nonumber \\
& \times \Bigg(
\ii \, (k_1-k_2) \, \NOthree{\psi_1}{\psi_2}{\psi_3}+
\bigg( (k_2+2)\, \NO{J_i^{(1)}}{\psi_i}
-\big(1 \leftrightarrow 2 \big) \bigg) \Bigg)
 , \\
H &=
-\frac{1}{\left(
3\, k_1 \, k_2 \, 
(k_1+2) (k_2+2) (k_1+k_2+4) (k_1+k_2+6)\right)^{1/2}} \times \nonumber \\
& \times \Bigg(
3\, \ii \,  k_1 \, k_2 \,(k_1+k_2+4) \, 
\NOthree{\psi_1}{\psi_2}{\psi_3}+
\bigg(2\, k_2 \, (2\, k_1+k_2+6)\, \NO{J_i^{(1)}}{\psi_i}+
\big(1 \leftrightarrow 2 \big) \bigg) \Bigg)
 , \\
M &=
\frac{\cpl_{k_1,k_2}}{4\,(2\, k_1+k_2+6)(k_1+ 2\, k_2+6)}
\Bigg(
6 \, (k_1 \, k_2 -2 \, k_1-2 \, k_2 -12) 
\, \NO{J_i^{(1)}}{J_i^{(2)}}
\Bigg. \nonumber \\
& -3 \, k_1 \, k_2 \, (k_1+k_2+4) \, \NO{J_i^{(3)}}{J_i^{(3)}}
+ \bigg( \big(-\frac{6 \, k_2 \, (k_2+2) (2\, k_1 +k_2 +6)}{k_1-k_2}
 \, \NO{J_i^{(1)}}{J_i^{(1)}} \big. \bigg. \nonumber \\
& \Bigg.
\bigg. \big. + \frac{3\, k_2 \, (k_1+k_2+4)
(3 \, k_1 \, k_2 +10 \, k_1+2\, k_2 +12)}
{k_1-k_2}
\, \NO{J_i^{(1)}}{J_i^{(3)}} 
\big) 
+\big(1 \leftrightarrow 2 \big) \bigg)
\Bigg)
 , \\
W &=  
{\left( \frac{c_{k_1, k_2}}
{9 \, k_1 \, k_2 \, (k_1+2)(k_2+2)(k_1+k_2+4) (k_1+k_2+6)}
\right)}^{1/2}
\times \nonumber \\
& \times
\Bigg( \half k_1 \, k_2 (k_1+k_2+4) 
\, \NO{J_i^{(3)}}{J_i^{(3)}}  
+2 (k_1 \, k_2 +4\, k_1 +4\, k_2+12) 
\, \NO{J_i^{(1)}}{J_i^{(2)}} \Bigg. \nonumber \\
& \Bigg. + \bigg(  \big(
- k_2 (k_2+2) \, \NO{J_i^{(1)}}{J_i^{(1)}} 
-2\, k_2 (k_1+k_2+4) \, \NO{J_i^{(1)}}{J_i^{(3)}} 
\big) + 
\big( 1 \leftrightarrow 2 \big) \bigg) \Bigg) 
 , \\
U &= 
{\left(\frac{2 \, c_{k_1, k_2}}
{9 \, k_1 \, k_2  \, (k_1+k_2+6)}
\right)}^{1/2} \times \nonumber \\
& \times
\left(
-6\, \ii \, 
{:} \left|
\barray{ccc}
J_1^{(1)} & J_2^{(1)} & J_3^{(1)} \\
J_1^{(2)} & J_2^{(2)} & J_3^{(2)} \\
\psi_1 & \psi_2 & \psi_3 \\
\earray
\right| {:}+
\bigg( \big(k_2 \, \NO{\dd J_i^{(1)}}{\psi_i}
-2\, k_2 \, \NO{J_i^{(1)}}{\dd \psi_i} \big)
+ \big( 1 \leftrightarrow 2 \big) \bigg)
\right) .
\end{align}
}

\subsection{\sww minimal models}

\setcounter{equation}{0}

\label{list rep}

%%%%%%%%%%%%%%%%%%%%%%%%%%%%%%%%%%%%%%%%%%%%%%%%%%%%%%%%%%%%%%%%

Here we present the complete list of 
unitary \hwr s of the \sww minimal models.
The central charge $c$ and the coupling $\cpl$
of the $(k_1, k_2)$ minimal model are given in 
(\ref{c cpl c}, \ref{c cpl cpl}).
The list of NS and Ramond sector representations can be 
presented in the form of three--dimensional table
with indices $s_1$, $s_2$, $s_3$:
\begin{equation}   \label{s range}
\begin{aligned}
s_1&= 1,2,\ldots,k_1+1, \\
s_2&= 1,2,\ldots,k_2+1, \\
s_3&= 1,2,\ldots,k_1+k_2+3 .
\end{aligned}
\end{equation}
The representation ${(s_1,s_2,s_3)}$ is of 
 NS or Ramond type
depending on $s_1+s_2+s_3$ is odd or even respectively.
The \hwr\ is labeled by 3 weights 
$\dm{1}$, $\dm{2}$, $\dm{3}$, the conformal dimensions
with respect to the three \none \sprc subalgebras.
Their values are taken from the correspondent
\none Kac tables (\ref{eq:N=1_dimensions}):
\begin{equation}
\dm{1}=d^{k_1}_{m_1,n_1}\, , \qquad 
\dm{2}=d^{k_2}_{m_2,n_2}\, , \qquad 
\dm{3}=d^{k_1+k_2}_{m_3,n_3}\, ,
\end{equation}
 where the indices are connected to $s_1$, $s_2$, $s_3$ by
\begin{equation}                      \label{m n min_mod}
\begin{aligned}
n_1&=s_1 \, ,\\ n_2&=s_2 \, ,\\ m_3&=s_3 \, ,
\end{aligned}
\qquad
\begin{aligned}
m_1&=s_1-Y_1 +Y_2+Y_3 \pm r  ,\\
m_2&=s_2+Y_1 -Y_2+Y_3 \pm r  ,\\
n_3&=s_3+Y_1 +Y_2-Y_3 \pm r  .
\end{aligned}
\end{equation}
$Y_1$, $Y_2$, $Y_3$ are values of the $Y_{a,b}(x)$ function: 
\begin{equation}
\begin{aligned}
Y_1&=Y_{2,2 k_2+2}\,(s_1-s_2-s_3+1), \\
Y_2&=Y_{2,2 k_1+2}\,(s_2-s_3-s_1+1), \\
Y_3&=Y_{2,2 \min(k_1,k_2)+2}\,(s_3-s_1-s_2+1).
\end{aligned}
\end{equation}
We define the function $Y_{a,b}(x)$  by its graph:
\begin{center}
%{\centering
\includegraphics[width=0.95 \textwidth]{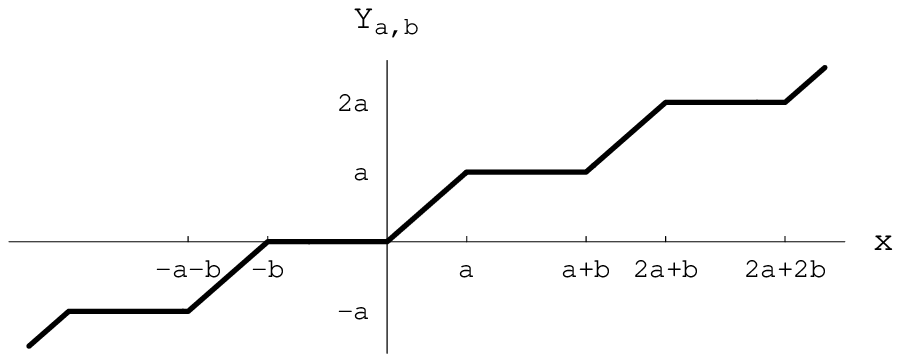}
%}
\end{center}
The number $r$ in (\ref{m n min_mod})
can be 0 or 1. It is 0 in the NS sector
($s_1+s_2+s_3$ odd). In the Ramond sector it is given by
\begin{align}
r&=1-\text{sgn}
\Big(
Y'_{2,2 k_2+2}\,(s_1-s_2-s_3+1) \Big. \nonumber \\
& \quad \Big. +Y'_{2,2 k_1+2}\,(s_2-s_3-s_1+1)
+Y'_{2,2 \min(k_1,k_2)+2}\,(s_3-s_1-s_2+1)
\Big),
\end{align}
where $Y'_{a,b}(x)$ is a derivative of $Y_{a,b}(x)$
with respect to $x$. 
The $Y'$ function is not continuous, but the values
in the points of discontinuity are not important.
$r$ distinguishes between
one ($r=0$) and two--dimensional ($r=1$) Ramond representations.

The conformal dimension $h$ (the eigenvalue of $L_0$
operator) is calculated from  $\dm{1}$, $\dm{2}$, $\dm{3}$
weights by
\begin{equation}   \label{h ddd}
 h= \frac{1}{2}\,\Big( - 
   \dm{3}\left({k_1} +{k_2} + 2 \right) 
       + \dm{1}
      \left({k_1} + 4 \right)  + 
    \dm{2} \left({k_2} + 4 \right)
\Big)
\end{equation}
in the case of one--dimensional (NS or Ramond) representation.
In the case of two--dimensional Ramond representation
the $\dm{1}$, $\dm{2}$, $\dm{3}$  in (\ref{h ddd})
should be substituted by half of the sum of the 
correspondent \none Ramond dimensions (\ref{h ddd R}).

The following representations are identical:
\begin{equation}
  \label{eq:rep ident}
  {(s_1, s_2, s_3)}={(k_1+2-s_1, k_2+2-s_2, k_1+k_2+4-s_3)}.
\end{equation}
%The ${(s_1, s_2, s_3)}$ representation is identical
%to the ${(k_1+2-s_1, k_2+2-s_2, k_1+k_2+4-s_3)}$ representation.
There are $[ \frac{(k_1+1)(k_2+1)(k_1+k_2+3)+1}{4} ]$
NS representations and the same number of Ramond 
representations.

The $k_1=k_2=k$ minimal models contain two additional
twisted sectors: tw1 and tw2.
The list of tw1 and tw2 representations can be
arranged in the two--dimensional table with indices
$t_1$ and $t_2$:
\begin{equation}   \label{t range}
\begin{aligned}
t_1&= 1,2,\ldots,k+2, \\
t_2&= 1,2,\ldots,k+1.
\end{aligned}
\end{equation}
The $t_1+t_2$ even entries are of tw2 type and
the $t_1+t_2$ odd entries are of tw1 type.
The representations in the twisted sectors are labeled by
the conformal dimension $h$ and the weight of the
third \none subalgebra $\dm{3}$:
\begin{equation}             \label{list tw reps}
\barray{rclcrcl}
h&=& \dis
\frac{|t_1-t_2|}{4}+
\frac{t_2^2-t_1^2+\delta}{8\,(k+2)} \, , &\, &
\delta&=&\left\{ 
\begin{array}{ll}
k-1, \phantom{t_2-1} %&  {(t_1+t_2) \bmod 2}=0 
& \quad\text{tw1},\\
3\,k/2, %&   {(t_1+t_2) \bmod 2}=1 
& \quad\text{tw2};
\end{array}
\right.
\\[16pt]
\dm{3}&= & \dis
\dm{2\,k}_{m,n}\, ,\quad 
\begin{aligned}
m&=t_1, \\
n&=t_1+\text{sgn}(t_2-t_1) \pm r,
\end{aligned}
 &\quad &
r&=&\left\{ 
\begin{array}{ll}
0, %&  {(t_1+t_2) \bmod 2}=0 
& \quad\text{tw1},\\
\text{sgn}(t_2-t_1), %&   {(t_1+t_2) \bmod 2}=1 
& \quad\text{tw2}.
\end{array}
\right.
\earray
\end{equation}
Again $r$ distinguishes between
one ($r=0$) and two--dimensional ($r=1$) tw2 representations.
There are $(k+1)(k+2)/2$ tw1 representations and the same number of 
tw2 representations.

We want to illustrate the formulas of the present appendix 
by some explicit examples. 
The simplest model is the $k_1=k_2=1$
($c=3/2, \cpl=0$) model, discussed
in section~\ref{k1_k2_1}. Its \hwr s
are presented in table~\ref{table_k_1=k_2=1}.
We use the $\kt{h}{\dm{1}}{\dm{2}}{\dm{3}}$ notation
for the NS and Ramond representations and the
$\kttw{h}{\dm{3}}$ notation for the twisted sectors.
One of the four NS/Ramond weights is dependent
on other three and is presented for convenience only.
The Ramond and tw1 sectors are {\it slanted}.
The $h=9/16$ Ramond representation is two-dimensional.

\begin{table}[!t]
%\centering
$$
\!\!\!\!\!
\setlength\arraycolsep{3pt}
\begin{array}{c}
\phantom{s_3 s_3 s_3} s_2 \rightarrow \hfill\\[2pt]
{\begin{array}{c}
s_3\\
\downarrow
\vspace{6cm}
\end{array}
}
{\setlength\arraycolsep{2pt}
 \begin{array}{|cc|}
\hline
% \rule{0cm}{30pt}
 \kt{0}{0}{0}{0} & \ktR{\frac{7}{16}}{\frac{3}{80}}{\frac{7}{16}
    }{\frac{3}{8}} \\[7pt]
 \ktR{\frac{1}{16}}{\frac{3}{80}}{\frac{3}{80}}{\frac{1}{16}} 
& \kt{\frac{1}{8}}{0}{\frac{1}{10}}{\frac{1}{16}}  \\[7pt]
  \kt{\frac{1}{6}}{\frac{1}{10}}{\frac{1}{10}}{\frac{1}{6}} & \ktR{\frac{5}{48
    }}{\frac{3}{80}}{\frac{3}{80}}{\frac{1}{24}}  \\[7pt]
  \ktR{\frac{9}{16}}{{
    \frac{3}{80} \atop \frac{7}{16}} }{{ \frac{3}{80} \atop \frac{7}{16}} }{
  {  \frac{1}{16} \atop \frac{9}{16}} } & \kt{\frac{1}{8}}{\frac{1}{10}}{0}{
\frac{1}{16}}  \\[7pt]
  \kt{\frac{1}{2}}{\frac{1}{10}}{\frac{1}{10}}{0
    } & \ktR{\frac{7}{16}}{\frac{7}{16}}{\frac{3}{80}}{\frac{3}{8}}\\[17pt]
\hline
\end{array}
}
\\ \\[-8pt]
\phantom{s_3 s_3} s_1=1
%\\[-15pt]
\end{array}
\,%\quad
\begin{array}{c}
\phantom{s_3 s_3 s_3} s_2 \rightarrow \hfill\\[2pt]
{\begin{array}{c}
s_3\\
\downarrow
\vspace{6cm}
\end{array}
}
{\setlength\arraycolsep{2pt}
\begin{array}{|cc|}
\hline
 \ktR{\frac{7}{16}}{\frac{7}{16}}{\frac{3}{80}}{\frac{3}{8}} 
& \kt{\frac{1}{2}}{\frac{1}{10}}{\frac{1}{10}}{0}  \\[7pt]
  \kt{\frac{1}{8}}{\frac{1}{10}}{0}{\frac{1}{16}} & 
 \ktR{\frac{9}{16}}{{
    \frac{3}{80} \atop \frac{7}{16}} }{{ \frac{3}{80} \atop \frac{7}{16}} }{
  {  \frac{1}{16} \atop \frac{9}{16}} }  \\[7pt]
  \ktR{\frac{5}{48}}{\frac{3}{80}}{\frac{3}{80}}{\frac{1}{24}} 
& \kt{\frac{1}{6}}{\frac{1}{10}}{\frac{1}{10
    }}{\frac{1}{6}}  \\[7pt]
  \kt{\frac{1}{8}}{0}{\frac{1}{10}}{\frac{1}{16}
    } & \ktR{\frac{1}{16}}{\frac{3}{80}}{\frac{3}{80}}{\frac{1}{16}
    }  \\[7pt]
  \ktR{\frac{7}{16}}{\frac{3}{80}}{\frac{7}{16}}{\frac{3}{8}
    } & \kt{0}{0}{0}{0} \\[17pt]
\hline
\end{array}
}
\\ \\[-8pt]
\phantom{s_3 s_3} s_1=2
%\\[-15pt]
\end{array}
\quad
%$$
%$$
\begin{array}{c}
\phantom{s_3 s_3 s_3} t_2 \rightarrow \hfill \\[2pt]
{\begin{array}{c}
t_1\\
\downarrow
\vspace{3cm}
\end{array}
}
{\setlength\arraycolsep{2pt}
\begin{array}{|cc|}
\hline
 \kttw{\frac{1}{16}}{0} & \kttwI{\frac{3}{8}}{\frac{3}{8}
   } \\[7pt] \kttwI{\frac{1}{8}}{\frac{1}{16}} & \kttw{\frac{1}{16}}{\frac{1}{1
   6}} \\[7pt] \kttw{\frac{11}{48}}{\frac{1}{6}} & \kttwI{\frac{1}{24}}{\frac{1
   }{24}}   
\\[17pt]
\hline
\end{array}
}
\\ \\[-8pt]
\phantom{ab_1} %s_1=1
\text{twisted}
\end{array}
%\quad
$$
\\[-15pt]
\phantom{aaaaaaaaaaaaaaaaaaaaaa}
NS and Ramond
\caption{The $k_1=k_2=1$
($c=3/2, \cpl=0$)  minimal model.}
\label{table_k_1=k_2=1}
\end{table}

The second example is the $k_1=2, k_2=3$ 
($c=37/15, \cpl=-182/(405 \sqrt{11})$) minimal model.
Since the list of representations is too long,
we reproduce only the conformal dimensions $h$
of the \hwr s (table~\ref{k_1=2_k_2=3}).

\begin{table}[t]
$$
\begin{array}{c}
\phantom{s_3 s_3 s_3 s_3} s_2 \rightarrow \hfill\\[2pt]
{\begin{array}{c}
s_3\\
\downarrow
\vspace{4.2cm}
\end{array}
}
\setlength\arraycolsep{2pt}
 \begin{array}{|cccc|}
\hline
\rule{0cm}{15pt}
\kth{0} & \ktRh{\frac{27}{80}} & \kth{\frac{9}{10}} & \ktRh{\frac{31}{16}} \\[7pt] \ktRh{
    \frac{5}{48}} & \kth{\frac{1}{15}} & \ktRh{\frac{121}{240}} & \kth{\frac{7}{6}} \\[7pt] \kth{\frac{
    5}{18}} & \ktRh{\frac{83}{720}} & \kth{\frac{8}{45}} & \ktRh{\frac{103}{144}} \\[7pt] \ktRh{\frac{
    37}{48}} & \kth{\frac{7}{30}} & \ktRh{\frac{41}{240}} & \kth{\frac{1}{3}} \\[7pt] \kth{\frac{5}{6}
    } & \ktRh{\frac{161}{240}} & \kth{\frac{7}{30}} & \ktRh{\frac{13}{48}} \\[7pt] \ktRh{\frac{175}{14
    4}} & \kth{\frac{61}{90}} & \ktRh{\frac{443}{720}} & \kth{\frac{5}{18}} \\[7pt] \kth{\frac{7}{6}
    } & \ktRh{\frac{241}{240}} & \kth{\frac{17}{30}} & \ktRh{\frac{29}{48}} \\[7pt] \ktRh{\frac{23}{16
    }} & \kth{\frac{9}{10}} & \ktRh{\frac{67}{80}} & \kth{1} \\[7pt] 
\hline
\end{array}
\\ \\[-8pt]
\phantom{s_3 s_3} s_1=1
\end{array}
\quad
\begin{array}{c}
\phantom{s_3 s_3 s_3 s_3} s_2 \rightarrow \hfill\\[2pt]
{\begin{array}{c}
s_3\\
\downarrow
\vspace{4.2cm}
\end{array}
}
\setlength\arraycolsep{2pt}
 \begin{array}{|cccc|}
\hline
\rule{0cm}{15pt}
  \ktRh{\frac{3}{8}} & \kth{\frac{27}{80}} & \ktRh{\frac{31}{40}} & \kth{\frac{23}{16}
    } \\[7pt] \kth{\frac{5}{48}} & \ktRh{\frac{53}{120}} & \kth{\frac{121}{240}} & \ktRh{\frac{25}{24}
    } \\[7pt] \ktRh{\frac{11}{72}} & \kth{\frac{83}{720}} & \ktRh{\frac{199}{360}} & \kth{\frac{103}{1
    44}} \\[7pt] \kth{\frac{13}{48}} & \ktRh{\frac{13}{120}} & \kth{\frac{41}{240}} & \ktRh{\frac{17}{
    24}} \\[7pt] \ktRh{\frac{17}{24}} & \kth{\frac{41}{240}} & \ktRh{\frac{13}{120}} & \kth{\frac{13}{
    48}} \\[7pt] \kth{\frac{103}{144}} & \ktRh{\frac{199}{360}} & \kth{\frac{83}{720}} & \ktRh{\frac{1
    1}{72}} \\[7pt] \ktRh{\frac{25}{24}} & \kth{\frac{121}{240}} & \ktRh{\frac{53}{120}} & \kth{\frac{
    5}{48}} \\[7pt] \kth{\frac{23}{16}} & \ktRh{\frac{31}{40}} & \kth{\frac{27}{80}} & \ktRh{\frac{3}{
    8}} \\[7pt]
  \hline
\end{array}
\\ \\[-8pt]
\phantom{s_3 s_3} s_1=2
\end{array}
\quad
\begin{array}{c}
\phantom{s_3 s_3 s_3 s_3} s_2 \rightarrow \hfill\\[2pt]
{\begin{array}{c}
s_3\\
\downarrow
\vspace{4.2cm}
\end{array}
}
\setlength\arraycolsep{2pt}
 \begin{array}{|cccc|}
\hline
\rule{0cm}{15pt}
\kth{1} & \ktRh{\frac{67}{80}} & \kth{\frac{9}{10}} & \ktRh{\frac{23}{16}
    } \\[7pt] \ktRh{\frac{29}{48}} & \kth{\frac{17}{30}} & \ktRh{\frac{241}{240}} & \kth{\frac{7}{6}
    } \\[7pt] \kth{\frac{5}{18}} & \ktRh{\frac{443}{720}} & \kth{\frac{61}{90}} & \ktRh{\frac{175}{144
    }} \\[7pt] \ktRh{\frac{13}{48}} & \kth{\frac{7}{30}} & \ktRh{\frac{161}{240}} & \kth{\frac{5}{6}
    } \\[7pt] \kth{\frac{1}{3}} & \ktRh{\frac{41}{240}} & \kth{\frac{7}{30}} & \ktRh{\frac{37}{48}
    } \\[7pt] \ktRh{\frac{103}{144}} & \kth{\frac{8}{45}} & \ktRh{\frac{83}{720}} & \kth{\frac{5}{18}
    } \\[7pt] \kth{\frac{7}{6}} & \ktRh{\frac{121}{240}} & \kth{\frac{1}{15}} & \ktRh{\frac{5}{48}
    } \\[7pt] \ktRh{\frac{31}{16}} & \kth{\frac{9}{10}} &
  \ktRh{\frac{27}{80}} & \kth{0} \\[7pt]  
\hline
\end{array}
\\ \\[-8pt]
\phantom{s_3 s_3} s_1=3
\end{array}
$$
\caption{The $k_1=2, k_2=3$  minimal model.}
\label{k_1=2_k_2=3}
\end{table}

%\newpage

%%%%%%%%%%%%%%%%%%%%%%%%%%%%%%%%%%%%%%%%%%%%%%%%%%%%%%%%%%%%%%%%

\subsection{Mode expansions of normal ordered products}

\setcounter{equation}{0}

\label{NOexp}

%%%%%%%%%%%%%%%%%%%%%%%%%%%%%%%%%%%%%%%%%%%%%%%%%%%%%%%%%%%%%%%%

Here we derive the formula for the
mode expansion of normal ordered product of operators
in various sectors.
The normal ordered product $\NO{P}{Q}$ is defined as the zero order term 
in \ope :
\begin{equation}
P(z) \, Q(w)=
%\mathrm{singular\ terms}
\sum_{k=1}^N \frac{R^{(k)}(w)}{(z-w)^k}
+\NO{P}{Q}(w)+O(z-w) \, .
\end{equation}
The well known formula for the mode expansion of $\NO{P}{Q}$
\begin{equation}                       \label{NOexpansion}
  \NO{P}{Q}_n =\sum_{m \le -\Delta_P} P_m Q_{n-m}+
(-1)^{PQ} \! \! \! \! \! \sum_{m \ge -\Delta_P +1} Q_{n-m} P_m \, ,
\end{equation}
is valid only if $m$ has the same modding as $\Delta_P$,
i.e. $m$ runs on integer  or  half integer numbers
depending on 
the spin of $P$ is integer or  half integer respectively. 
 So in the NS sector
the expansion (\ref{NOexpansion}) works. In the case
of Ramond or twisted sectors the formula should be
modified.

The idea of the following calculation is taken 
from ref.\cite{Odake:1988bh} (section 3),
where the mode expansion of $\NO{G^+}{G^-}$
was obtained using the same method.
($G^+$ and $G^-$ are the supersymmetry generators
of the $N=2$ superconformal algebra.) 
Let's calculate the integral:
\begin{equation}  \label{NOintegral}
\oint_0  \! \mathrm{d} w \, w^{n+\Delta_P+\Delta_Q-1}
\oint_w  \! \mathrm{d} z \, \frac{1}{z-w} \, \,
z^{\epsilon} \, P(z) \, Q(w) \, w^{-\epsilon} \, ,
\end{equation}
where the first integration is around $w$ and the second
is around $0$.
The integral (\ref{NOintegral}) is equal to the $n$--mode
\begin{equation}
\NO{P}{Q}_n + 
\sum_{k=1}^N { \epsilon \choose k} \, R^{(k)}_n \, .
\end{equation}

The integration contour in (\ref{NOintegral}) 
can be transformed to 
\begin{equation}
{\oint \!  \! \oint \!  \!}_{z>w} \mathrm{d} z \, \mathrm{d} w-
{\oint \!  \! \oint \!  \!}_{w>z} \mathrm{d} w \, \mathrm{d} z \, .
\end{equation}
The $z^{\epsilon}$ term in the integration function 
was introduced to compensate the phase change
of $P(z)$ around $z=0$.

Expanding $(z-w)^{-1}$ and integrating one gets
\begin{equation}   \label{NOepsilon}
\NO{P}{Q}_n = 
- \sum_{k=1}^N { \epsilon \choose k} \, R^{(k)}_n+
\sum_{m \le -\Delta_P + \epsilon} P_m Q_{n-m}+
(-1)^{PQ} \! \! \! \! \! \sum_{m \ge -\Delta_P +1+\epsilon} Q_{n-m} P_m \, ,
\end{equation}
where $m$ runs on $\mathbb{Z}-\Delta_P + \epsilon$.
$\epsilon$ is usually chosen to be $0$ or $1/2$ 
to produce the correct modding
for operator $P$. In the case $\epsilon=0$
we get back the formula  (\ref{NOexpansion}) as expected.
Note that the formula (\ref{NOepsilon}) is valid
for any $\epsilon$ (not only 0 or $1/2$) consistent with
the chosen modding.

Another approach to the calculation of mode expansions
of composite operators is presented in \cite{Eholzer:1992pv}
(section 3) and in \cite{gn} (appendix C)
and leads to the same results.

%%%%%%%%%%%%%%%%%%%%%%%%%%%%%%%%%%%%%%%%%%%%%%%%%%%%%%%%%%%%%%%%

\end{document}